\documentclass[aps,pra,twocolumn,groupedaddress,floatfix]{revtex4}
\usepackage{amsfonts}
\usepackage{amsfonts}
\usepackage{amsmath}
\usepackage{amssymb}
\usepackage{epsfig}
\usepackage{graphicx}
\usepackage{subfigure}
\usepackage{float}
\usepackage{color}
\usepackage[colorlinks,
linkcolor=blue,       
anchorcolor=blue,  
urlcolor=blue,        
citecolor=blue,      
]{hyperref}
\usepackage{natbib}

\begin{document}

\title{Optimized nonadiabatic holonomic quantum computation based on F\"orster resonance in Rydberg atoms}

\author{Shuai Liu$^{1,2}$, Jun-Hui Shen$^{3,1,}$\footnote{13848612@qq.com}, Ri-Hua Zheng$^{1,2}$, Yi-Hao Kang$^{1,2}$, Zhi-Cheng Shi$^{1,2}$, Jie Song$^{4}$, and Yan Xia$^{1,2,}$\footnote{xia-208@163.com}}
\address{$^1$Fujian Key Laboratory of Quantum Information and Quantum Optics (Fuzhou University), Fuzhou 350116, China\\
$^2$Department of Physics, Fuzhou University, Fuzhou 350116, China\\
$^3$School of Rail Transportation, Fujian Chuanzheng Communications College, Fuzhou 350007, China\\
$^4$Department of Physics, Harbin Institute of Technology, Harbin 150001, China}


\begin{abstract}

In this paper, we propose a scheme for implementing the nonadiabatic holonomic quantum computation (NHQC+) of two Rydberg atoms by using invariant-based reverse engineering (IBRE). The scheme is based on F\"orster resonance induced by strong dipole-dipole interaction between two Rydberg atoms, which provides a selective coupling mechanism to simply the dynamics of system. Moreover, for improving the fidelity of the scheme, the optimal control method is introduced to enhance the gate robustness against systematic errors. Numerical simulations show the scheme is robust against the random noise in control fields, the deviation of dipole-dipole interaction, the F\"orster defect, and the spontaneous emission of atoms. Therefore, the scheme may provide some useful perspectives for the realization of quantum computation with Rydberg atoms.

\end{abstract}

\maketitle

\section{Introduction}
The quantum computation, due to the characteristic of coherent superposition of quantum states, has shown many advantages in speeding up the processing of certain complex problems, such as factoring large integers and searching unsorted databases \cite{PhysRevLett.79.325,PhysRevA.64.022307,PhysRevLett.95.140501,Fan2018,Fan2019}. In quantum computing,
 the realization of  high-fidelity gates is necessary for fault-tolerant quantum computing. However, the gate fidelity is generally limited by control errors in gate parameters and decoherence induced by dissipation. To diminish the effects of control errors and decoherence, various proposals of quantum computation have been put forward. One promising proposal is
nonadiabtic holonomic quantum computation (NHQC) based on a non-Abelian geometric phases  \cite{sjoqvist2012non,PhysRevLett.109.170501,PhysRevA.99.052309,PhysRevA.101.062306,PhysRevA.101.012306,PhysRevA.98.032313,PhysRevA.95.032311,PhysRevA.98.052314,PhysRevLett.124.230503,Li2021}. On one hand, the NHQC is based on geometric phases determined by global properties of evolution paths. Therefore, it is insensitive to classical parameter fluctuation. On the other hand, the NHQC is free from the limitation of the adiabatic condition, thus the evolution process can be accelerated. Accordingly, this can reduce the influence of decoherence on the system and improve the fidelity of the gate. However, the NHQC requires strict conditions to be satisfied for the Hamiltonian at all times, which undoubtedly restricts the choice of parameters \cite{sjoqvist2012non,PhysRevLett.109.170501}. 

To overcome the difficulty, a flexible scheme \cite{PhysRevLett.123.100501} called as the NHQC+ has recently been proposed, which relaxes the requirement of parameters by selecting auxiliary basis. With the relaxation of the conditions, the NHQC+ approach can be
compatible with various optimal control methods including counteradiabatic driving (CD) \cite{Berry2009,PhysRevLett.105.123003,PhysRevLett.111.100502,Qi:20}, dynamical decoupling (DD) \cite{PhysRevLett.95.180501,PhysRevLett.106.240501,PhysRevLett.118.133202}, single-shot-shaped pulse (SSSP) \cite{PhysRevLett.111.050404,PhysRevA.97.062317}, invariant based reverse engineering (IBRE) \cite{Lewis1969,PhysRevA.83.062116}, etc., which further improves the robustness against systematic errors. To date, many NHQC+ schemes have been put forward in different physical systems, for example,  superconducting circuit \cite{Li2020fast,PhysRevLett.122.080501,Xu2020}, spin qubits \cite{PhysRevA.101.032322,PhysRevA.101.052302}, and Rydberg atoms  \cite{PhysRevA.102.042607,kang2020heralded,PhysRevResearch.2.043130,PhysRevApplied.14.034038}.

The Rydberg atom system is one promising candidate platform for physical implementation of quantum computing due to its long coherence time and strong interatomic interaction \cite{RevModPhys.82.2313,Urban2009,Gatan2009,PhysRevLett.114.113002,PhysRevA.95.022319,PhysRevA.97.042336,PhysRevA.98.032306,PhysRevA.99.032348,PhysRevA.97.032328,PhysRevA.101.012345,PhysRevA.102.012609,PhysRevA.101.012347,Shi2021,PhysRevA.103.052402,Yin:20}. In Rydberg atom systems, the most representative phenomenon is Rydberg blockade \cite{PhysRevLett.85.2208,PhysRevLett.87.037901}. That is, when more than one atom are excited in Rydberg states, strong dipole-dipole or van der Waals interactions between Rydberg atoms exist, which prevents neighboring Rydberg atoms simultaneously being excited to Rydberg states. In contrast to the Rydberg blockade, the Rydberg antiblockade \cite{PhysRevLett.98.023002,PhysRevLett.104.013001,PhysRevA.96.042335} is more selective for computational basis states. For a two-qubit system, the Rydberg antiblockade allows only one computational basis state to be excited to the doubly excited Rydberg state, while the remaining three computational states keep unchanged. 
Apart from the Rydberg blockade and Rydberg antiblockade, the interaction among neutral atoms can exhibit many peculiar phenomena, such as, Rydberg dressing \cite{PhysRevA.89.032334,PhysRevA.91.012337,PhysRevA.101.030301} and F\"orster resonance \cite{PhysRevA.98.052324,PhysRevA.97.032701}. 

Among these different phenomena, the Starked F\"orster resonance \cite{Frster1948} is an useful way to control the interaction between atoms, where two pairs of Rydberg states allow dipole transitions in between can be shifted into resonance by dc or microwave electric fields \cite{PhysRevLett.104.073003,Gorniaczyk2016}. Moreover, the coherent coupling at F\"orster resonance has been also experimentally demonstrated \cite{Ravets2014} and exploited to realize quantum computing \cite{PhysRevA.82.034307,PhysRevA.102.053118}. For example, Beterov \emph{et al.} proposed a scheme for two-qubit gates using double adiabatic passage of the Stark-tuned F\"orster resonances of Rydberg states \cite{PhysRevA.94.062307}. Huang \emph{et al}. proposed a scheme to implement the two-qubit controlled-Z gate via the Stark-tuned F\"orster interaction of Rydberg atom \cite{PhysRevA.98.042704}. In view of that we are led to ask if it is possible to implement NHQC+ using the F\"orster resonance. The answer is positive.

In this paper, we propose a scheme to realize NHQC+ based on F\"orster resonance in Rydberg atoms. The scheme has several advantages as follows: Firstly, inspired by Refs.~\cite{PhysRevA.98.062338,PhysRevA.102.053118}, we realize a selective coupling mechanism for specific initial state in the regime of Rydberg F\"orster resonance, which is utilized to simplify the dynamics of system. Therefore, the mechanical effect \cite{PhysRevLett.94.173001} and the possible ionization \cite{PhysRevLett.94.173001,PhysRevLett.98.023004,PhysRevA.76.054702} can be effectively suppressed because there have no the simultaneous excitation of Rydberg atoms. Secondly, the evolution is studied by IBRE, where the paths for NHQC+ are naturally constructed by eigenvectors of an dynamic invariant and the corresponding control fields are also reversely designed. Moreover, the scheme can be compatible with the zero-systematic-error-sensitivity optimal control method \cite{Ruschhaupt2012}, so that the robustness of the control fields against systematic errors is further enhanced. The performance of the scheme is estimated with the numerical simulations and the results indicate that the scheme is insensitive to the random noise in control fields, the deviation of dipole-dipole interaction and the F\"orster defect. Thirdly, the decoherence is also taken into account. Because the populations of partial Rydberg states are decoupled with the effective Hamiltonian and quantum information is coded on the atoms' ground states, the scheme is robust against the spontaneous emission of atoms. Therefore, the scheme may provide useful perspectives in the realization of high-fidelity quantum computation.

The outline of the paper is as follows. In \mbox{Sec.~\ref{II}}, we review the general theories for realizing NHQC+ with IBRE.
In Sec.~\ref{III}, we implement nonadiabatic holonomic quantum gates in the regime of F\"orster resonance by using IBRE. In Sec.~\ref{IV}, we analyze the influence of errors and decoherence on the fidelity of the quantum gate based on the NHQC+ (we call it as NHQC+ gate throughout the paper) via numerical simulations. Finally, conclusions are presented in Sec.~\ref{VI}.
\section{theoretical preparation }\label{II}
\subsection{Lewis-Riesenfeld invariant theory}
For the sake of elaborating the scheme more clearly, let us first review the Lewis-Riesenfeld theory in a nutshell \cite{Lewis1969}. We now consider a system with Hamiltonian $H(t)$. By introducing a Hermitian invariant operator $I(t)$ obeying the equation ($\hbar=1$)
\begin{eqnarray}\label{eq1}
	i\frac{\partial}{\partial t}I(t)-[H(t),I(t)]=0,
\end{eqnarray}
an arbitrary solution of the time-dependent Schr\"odinger equation $i\frac{\partial}{\partial t}|\Psi(t)\rangle=H(t)|\Psi(t)\rangle$ can be expressed by eigenvectors $\{|\vartheta_{k}\rangle\}$ of $I(t)$ as
\begin{eqnarray}\label{eq2}
	|\Psi(t)\rangle&=&\sum_{k}c_k|\psi_{k}(t)\rangle,\cr
	|\psi_{k}(t)\rangle&=&\exp[i\alpha_{k}(t)]|\vartheta_{k}(t)\rangle,
\end{eqnarray}
where $k=0, 1, ...$, $c_k=\langle \vartheta_{k}(0)|\Psi(0)\rangle$ are corresponding coefficients and the Lewis-Riesenfeld phases $\alpha_k(t)$ are defined as
\begin{eqnarray}
	\alpha_{k}(t)=\int_{0}^{t}\langle \vartheta_{k}(t')|i\frac{\partial}{\partial t'}-H(t')|\vartheta_{k}(t')\rangle dt'.
\end{eqnarray}
 With the help of the dynamical invariant $I(t)$, the Hamiltonian $H(t)$ can be inversely derived by Eq. (\ref{eq1}). Therefore, dynamical invariant $I(t)$ is helpful for analyzing the evolution of the system.

\subsection{Requirements of the NHQC+ approach}
In this section, we briefly review the requirement of the NHQC+ approach \cite{PhysRevLett.123.100501}. Compared to the previous NHQC schemes, the difference is that the Hamiltonian in the NHQC+ approach \cite{PhysRevLett.123.100501} has fewer restrictions \cite{sjoqvist2012non,PhysRevLett.109.170501}. To facilitate the description of the difference more clearly, we consider an $M$-dimensional quantum system with Hamiltonian $H(t)$. Then, a complete set of basis can be represented as $\{|\phi_{m}(t)\rangle, m=1, 2, ..., M\}$, where all the basis $|\phi_{m}\rangle$ follow the Schr\"odinger equation. Generally speaking, the NHQC is realized by using a $L$-dimensional subspace satisfying the cyclic evolution and parallel transport conditions \cite{sjoqvist2012non,PhysRevLett.109.170501}, i.e.,
\begin{eqnarray}\label{eq4}
	(i)&&\sum_{m=1}^{L}|\phi_m(T)\rangle\langle\phi_m(T)|=	\sum_{m=1}^{L}|\phi_m(0)\rangle\langle\phi_m(0)|,\cr
		(ii)&&\langle \phi_m(t)|H(t)|\phi_{m'}(t)\rangle=0,  (m,m'=1, ..., L).
\end{eqnarray}
The above two conditions (i) and (ii) ensure a cyclic evolution with only pure geometric phases being accumulated during the time interval $[0,T]$. In the NHQC case, the condition (ii) shown in Eq.~(\ref{eq4}) requires restrictions for all possible $m, m'$ and at every moment, which hinders the combination of NHQC and some optimal control methods \cite{PhysRevLett.123.100501}.

To remove these constraints, the NHQC+ approach suggests to find a set of auxiliary basis $\{|\tilde{\vartheta}_m(t)\rangle, m=1, 2, ..., L\}$, which satisfies cyclic condition, i.e., $|\tilde{\vartheta}_m(T)\rangle=|\tilde{\vartheta}_m(0)\rangle=|\phi_m(0)\rangle$.
According to the results of Ref.~\cite{PhysRevLett.123.100501}, the defined projector
 $\Pi_m(t)=|\tilde{\vartheta}_m(t)\rangle\langle \tilde{\vartheta}_m(t)|$ should satisfy von Neumann equation
 \begin{eqnarray}\label{eq10}
 	\frac{d}{dt}\Pi_m(t)=-i[H(t),\Pi_m(t)].
 \end{eqnarray}
Besides, for each $|\widetilde{\vartheta}_m(t)\rangle$, the dynamic phase acquired in the whole evolution process should vanish, i.e.,
 \begin{eqnarray}\label{eq6}
 	\theta_m(T)=-\int^{T}_{0}\langle \tilde{\vartheta}_m(t)|H(t)|\tilde{\vartheta}_m(t)\rangle dt=0,
 \end{eqnarray}
such that the evolution becomes purely geometric
as
\begin{eqnarray}
	U(T,0)=\sum_{m}e^{i\Theta_m(T)}|\tilde{\vartheta}_m(T)\rangle\langle \tilde{\vartheta}_m(T)|,
\end{eqnarray}
with the geometric phase
\begin{eqnarray}\label{gphase}
 \Theta_m(T)=i\int_{0}^{T}\langle\tilde{\vartheta}_m(t)| \dot{\tilde{\vartheta}}_m(t)\rangle dt.	
\end{eqnarray}
 Compared with the condition (ii) in Eq.~(\ref{eq4}), the condition in Eq.~(\ref{eq6}) shows that the Hamiltonian in the NHQC+ approach removes the constraints for $m\neq m'$ and the dynamical phase is only required to
vanish for the integral in the time interval $[0,T]$.
Thus, the NHQC+ approach may be compatible with different optimal control methods.

\section{Implementation of nonadiabatic holonomic quantum gates}\label{III}
\begin{figure}[htbp] \centering
	\includegraphics[width=8.6cm]{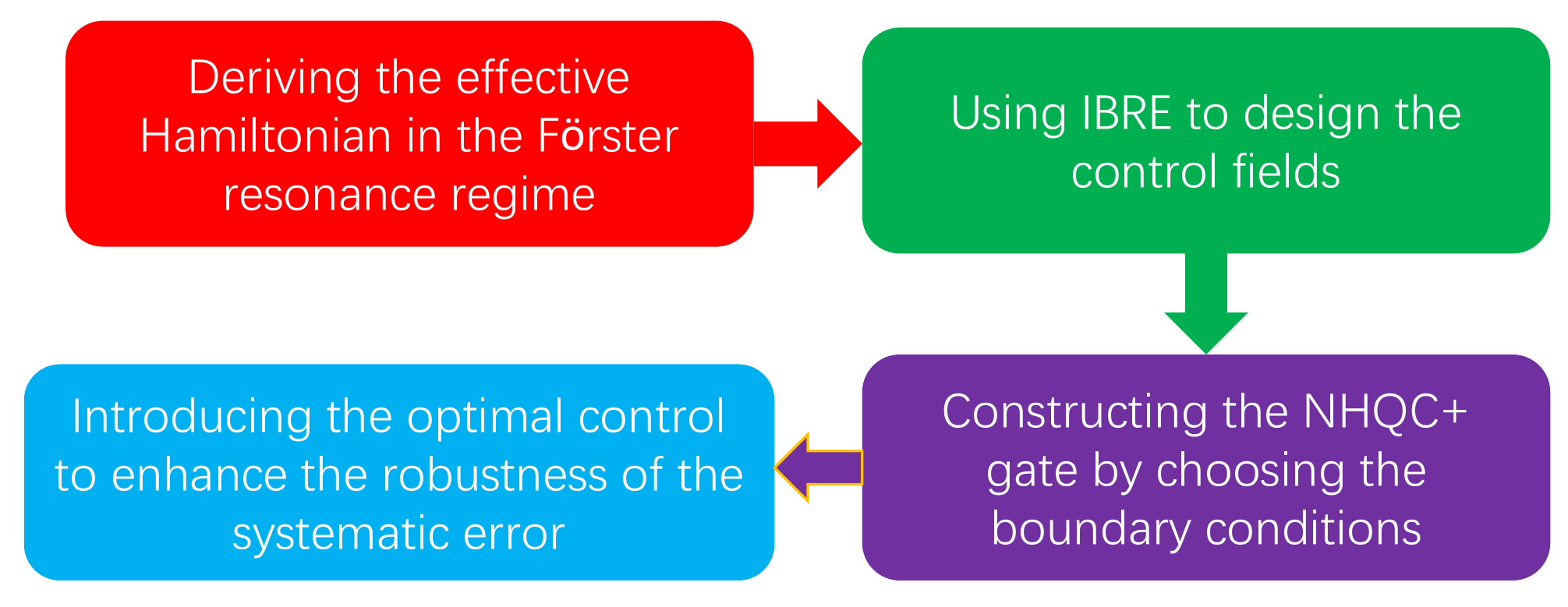}
	\caption{Schematic diagram for the implementation of optimized NHQC+ gates.}
	\label{fig0}
\end{figure}

In this section, we introduce the physical model, the design methods of control fields (IBRE), the optimal control method (zero-systematic-error-sensitivity optimal control) to implement the NHQC+ gate. In \mbox{Sec. \ref{III A}}, we give the physical model and derive the effective Hamiltonian in the F\"orster resonance regime. Based on the effective Hamiltonian in Sec.~\ref{III A}, we further combine IBRE with the optimal control method to realize the optimized NHQC+ gate in Sec.~\ref{III B}. Specially, a clear schematic diagram for the scheme is given in Fig.~\ref{fig0}.
\subsection{Physical model and effective Hamiltonian}\label{III A}

\begin{figure}[htbp] \centering
	\includegraphics[width=8cm]{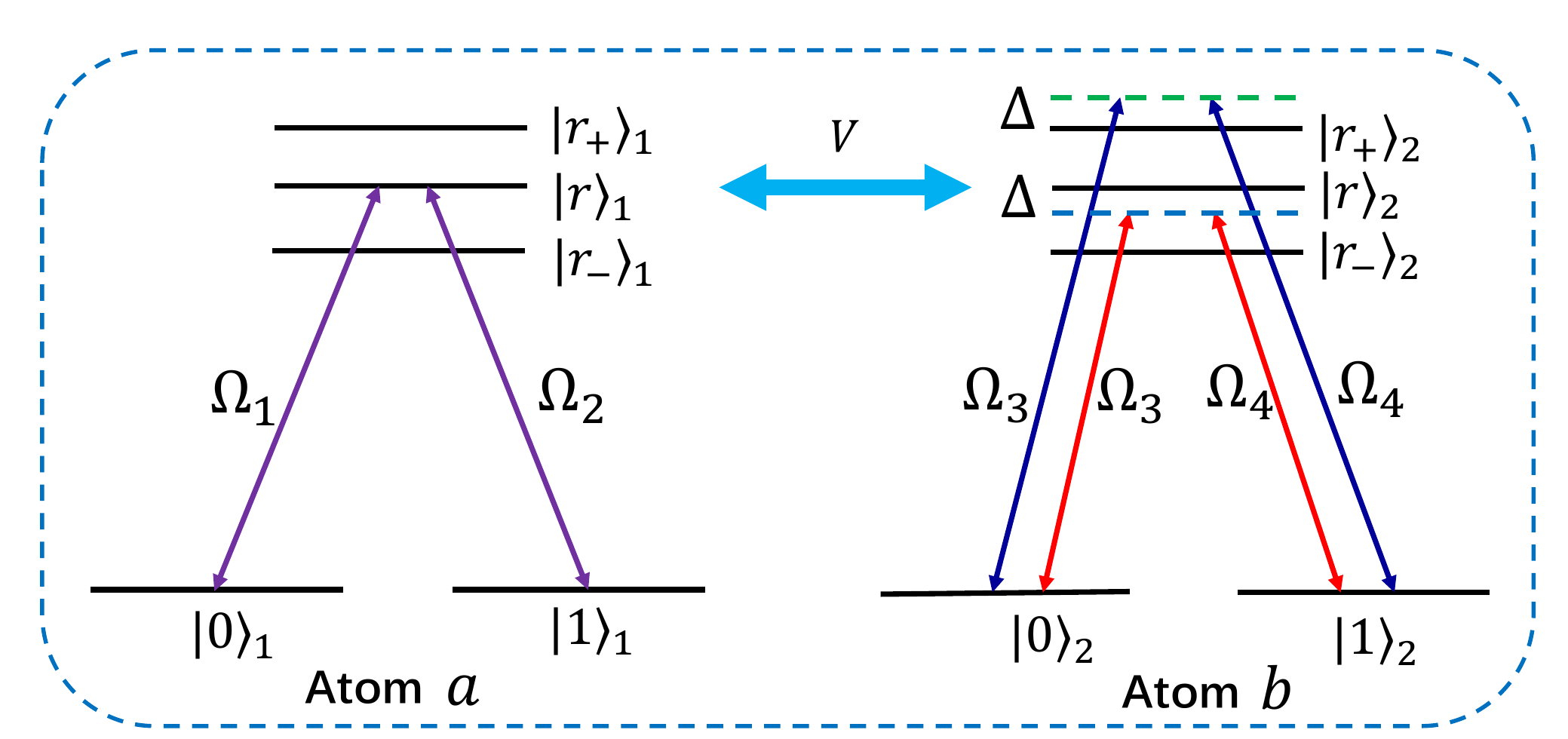}
	\caption{Level configurations and coherent couplings of atom $a$ and atom $b$ with a dipole-dipole interaction strength $V$.}
	\label{fig1}
\end{figure}

The model for implementing the nondiabatic holonomic quantum gates is shown in Fig.~\ref{fig1}.
We consider two identical Rubidium atoms individually trapped in optical tweezers. Each atom includes two ground states $|0\rangle=|5S_{1/2},F=1,m_{F}=0\rangle$ and $|1\rangle=|5S_{1/2},F=2,m_{F}=0\rangle$, and three Rydberg states $|r\rangle=|59D_{3/2},m_{j}=3/2\rangle$, $|r_{+}\rangle=|61P_{1/2},m_{j}=1/2\rangle$, and $|r_{-}\rangle=|57F_{5/2},m_{j}=5/2\rangle$ \cite{Walker2005}. The transition $|0\rangle_1(|1\rangle_1)\leftrightarrow|r\rangle_1$ for atom 1 is resonantly driven by control field with Rabi frequency $\Omega_{1}(t)$ ($\Omega_{2}(t)$). For the atom 2, the off-resonant transition $|0\rangle_{2}\leftrightarrow|r\rangle_{2} (|r_{+}\rangle_{2})$ is driven by control field of Rabi frequency $\Omega_{3}$ with red detuning $\Delta$ (blue detuning $\Delta$). Besides, the off-resonant transition $|1\rangle_{2}\leftrightarrow|r\rangle_2 (|r_{+}\rangle_2)$ is driven by control field of $\Omega_{4}$ with red detuning $\Delta$ (blue detuning $\Delta$).  When both atoms are excited to Rydberg states, the states $|rr\rangle$ and $(|r_{+}r_{-}\rangle+|r_{-}r_{+}\rangle)/\sqrt{2}$ are coupled by the dipolar interaction based on the F\"orster process with a small F\"orster defect, which the pair states $|rr\rangle$, $|r_{+}r_{-}\rangle$, and $|r_{-}r_{+}\rangle$ are almost degenerate \cite{Walker2005,PhysRevA.77.032723}. Noticing that the F\"orster defect
	can be eliminated by an external field \cite{Ravets2014}. Therefore, the resonant F\"orster interaction between two Rydberg atoms is described as
\begin{eqnarray}\label{eq99}
	H_{F}=V|rr\rangle(\langle r_{+}r_{-}|+\langle r_{-}r_{+}|)/\sqrt{2}+\mathrm{H.c.}
\end{eqnarray}
The Hamiltonian $H_F$ in Eq.~(\ref{eq99}) can be further diagonalized as $V(|\varpi_{+}\rangle\langle \varpi_{+}|+|\varpi_{-}\rangle\langle \varpi_{-}|)$ with $|\varpi_{\pm}\rangle=(|rr\rangle\pm|R\rangle)/\sqrt{2}$, where we define $|R\rangle=(|r_{+}r_{-}\rangle+|r_{-}r_{+}\rangle)/\sqrt{2}$.
To parameterize the Rabi frequencies of control fields  $\Omega_{1}(t)$,  $\Omega_{2}(t)$, $\Omega_{3}$ and $\Omega_{4}$,
we assume  $\Omega_1(t)=\Omega_a(t)\cos v_ae^{i\varphi_{a}(t)}$, $\Omega_{2}(t)=\Omega_{a}(t)\sin v_ae^{i\varphi_{a}(t)}$,
$\Omega_{3}=\Omega_{b}\cos v_b$, and  $\Omega_{4}=\Omega_{b}\sin v_b$.
In the interaction picture, the Hamiltonian $H_{I}$ of the system can be written as
\begin{eqnarray}\label{eq9}
H_{I}&=&H_{1}+H_{2}+H_{F},\cr\cr H_1&=&\Omega_{a}(t)e^{i\varphi_{a}(t)}|\xi_{+}\rangle_a\langle r|+\mathrm{H.c.},\cr\cr
H_2&=&\Omega_{b}e^{-i\Delta t}|r_{+}\rangle_{b}\langle \xi_{+}|+\Omega_{b}e^{i\Delta t}|r\rangle_b\langle\xi_{+}|+\mathrm{H.c.},\cr&&
\end{eqnarray}
where $|\xi_{+}\rangle_{l}=\cos v_l|0\rangle_l+\sin v_{l}|1\rangle_l$ ($l=a,b$).
\begin{figure}[hbpt]
	\centering
	\includegraphics[width=1\linewidth]{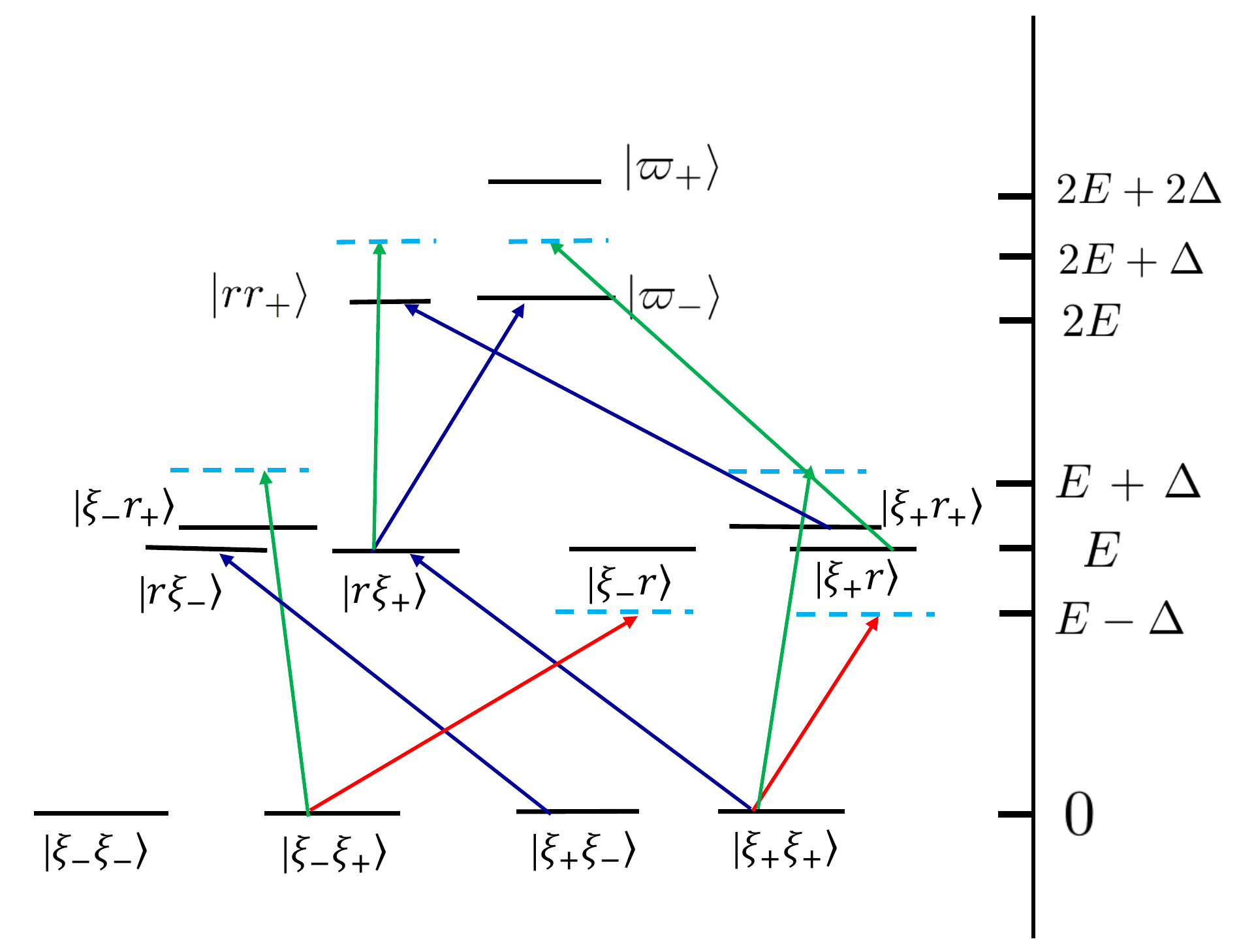}
	\caption{Setup for effective couplings of the two-atom collective ground states and the two-atom collective Rydberg state. The blue solid lines represent the resonance interaction between these two states. The red solid lines and green solid lines represent the interaction with red detuning $\Delta$ and blue detuning $\Delta$, respectively. $E$ denotes the energy of the single atom in the Rydberg state $|r\rangle$. }
	\label{fig2}
\end{figure}
Besides, we also define
 $|\xi_{-}\rangle_l=\cos v_l|1\rangle-\sin v_l|0\rangle_l$, such that $\{|\xi_{-}\rangle_{l},|\xi_{+}\rangle_{l}\}$ can form a complete orthogonal basis for atom $l$ instead of $\{|0\rangle_{l},|1\rangle_{l}\}$.

After considering the condition $\Delta=V$, the Hamiltonian $H_{I}$ in Eq.~(\ref{eq9}) is reformulated in a rotating frame with respect to $R=e^{iVt(|\varpi_{+}\rangle\langle\varpi_{+}|-|\varpi_{-}\rangle\langle\varpi_{-}|)}$,
\begin{eqnarray}\label{eq11}
H^{r}_{1}=RHR^{\dagger}+i\dot{R}R^{\dagger}=H_r+H_{h},
\end{eqnarray}
where the Hamiltonians $H_r$ and $H_{h}$ are respectively given by
\begin{eqnarray}\label{eq13}
H_r &=&\Omega_{a}(t)e^{i\varphi_{a}(t)}(|\xi_{+}\xi_{-}\rangle\langle r\xi_{-}|+|\xi_{+}\xi_{+}\rangle\langle r\xi_{+}|\cr
&&+|\xi_{+}r_{+}\rangle\langle rr_{+}|)
+\frac{\Omega_{b}}{\sqrt{2}}|r\xi_{+}\rangle\langle \varpi_{-}|+\mathrm{H.c.},\cr
H_{h}&=&\Omega_{b}e^{-i\Delta t}(|\xi_{-}\xi_{+}\rangle\langle \xi_{-}r|+|\xi_{-}r_{+}\rangle\langle \xi_{-}\xi_{+}|\cr\cr
&&+|\xi_{+}\xi_{+}\rangle\langle \xi_{+}r|
+|\xi_{+}r_{+}\rangle\langle \xi_{+}\xi_{+}|+|rr_{+}\rangle\langle r\xi_{+}|)\cr\cr
&&+\frac{\Omega_{a}(t)e^{i\varphi_{a}(t)}}{\sqrt{2}}e^{-i\Delta t}(|\varpi_{-}\rangle\langle \xi_{+}r|
+|\xi_{+}r\rangle\langle \varpi_{+}|)\cr
&&+\frac{\Omega_{b}}{\sqrt{2}}e^{-2i\Delta t}|r\xi_{+}\rangle\langle \varpi_{+}|+\mathrm{H.c.}
\end{eqnarray}

In the regime of the large detuning limit $\Delta\gg\Omega_{b}$, the Hamiltonian $H_{h}$ in Eq.~(\ref{eq13}) can be reduced as \cite{james2007effective}
\begin{eqnarray}\label{eq133}
H'_{h1}&=&\frac{\Omega_{b}^2}{\Delta}(|\xi_{-}r\rangle\langle \xi_{-}r|+|\xi_{+}r\rangle\langle \xi_{+}r|-|\xi_{-}r_{+}\rangle\langle \xi_{-}r_{+}|\cr\cr
&&+|\xi_{+}r\rangle\langle \xi_{+}r|-|\xi_{+}r_{+}\rangle\langle \xi_{+}r_{+}|+|r\xi_{+}\rangle\langle r\xi_{+}|\cr\cr
&&-|rr_{+}\rangle\langle rr_{+}|)+\frac{\Omega_{b}^2}{4\Delta}(|\varpi_{+}\rangle\langle \varpi_{+}|-|r\xi_{+}\rangle\langle r\xi_{+}|)\cr\cr&&+\frac{\Omega_{a}^2}{2\Delta}(|\varpi_{+}\rangle\langle \varpi_{+}|-|\varpi_{-}\rangle\langle \varpi_{-}|),\cr
H'_{h2}&=&-\frac{\Omega_{b}\Omega_{a}(t)e^{i\varphi_{a}(t)}}{\sqrt{2}\Delta}|\varpi_{-}\rangle\langle \xi_{+}\xi_{+}|+\mathrm{H.c.},\cr
&&
\end{eqnarray}
where, the term $H'_{h1}$ can be identified with Stark shifts associated with the control fields and the term $H'_{h2}$ represents the Raman transition $|\xi_{+}\xi_{+}\rangle\leftrightarrow |\varpi_{-}\rangle$.
If the system is initial in subspace $\mathcal{S}=\{|\xi_{-}\xi_{-}\rangle,|\xi_{-}\xi_{+}\rangle,|\xi_{+}\xi_{-}\rangle,|\xi_{+}\xi_{+}\rangle\}$, by neglecting decoupled terms, the full Hamiltonian $H_{1}^{r}$ in Eq.~(\ref{eq11}) can be simplified as
\begin{eqnarray}\label{eq14}
H^{r}_{2}&=&\frac{3\Omega^{2}_{b}}{4\Delta}|r\xi_{+}\rangle\langle r\xi_{+}|+\Omega_a(t)e^{i\varphi_{a}(t)}(|\xi_{+}\xi_{-}\rangle\langle r\xi_{-}|\cr
&&+|\xi_{+}\xi_{+}\rangle\langle r\xi_{+}|)
+\frac{\Omega_b}{\sqrt{2}}|r\xi_{+}\rangle\langle \varpi_{-}|+\mathrm{H.c.}
\end{eqnarray}
As shown in Fig.~\ref{fig2}, only the transitions $|\xi_{+}\xi_{-}\rangle\leftrightarrow|r\xi_{-}\rangle$, $|\xi_{+}\xi_{+}\rangle\leftrightarrow|r\xi_{+}\rangle$, and $|r\xi_{+}\rangle\leftrightarrow|\varpi_{-}\rangle$ are allowed in the regime of the large detuning limit $\Delta\gg\Omega_b$.

In order to achieve a selective coupling interaction, we further suppress the transition $|\xi_{+}\xi_{+}\rangle\leftrightarrow|r\xi_{+}\rangle$ by diagonalizing the Hamiltonian $\frac{\Omega_b}{\sqrt{2}}|r\xi_{+}\rangle\langle \varpi_{-}|+\mathrm{H.c.}$ in Eq.~(\ref{eq14}). That is, the term $\frac{\Omega_b}{\sqrt{2}}|r\xi_{+}\rangle\langle \varpi_{-}|+\mathrm{H.c.}$ in Eq.~(\ref{eq14}) is transformed to the form $\frac{\Omega_b}{\sqrt{2}}(|E_+\rangle\langle E_+|-|E_-\rangle\langle E_-|)$ with $|E_{\pm}\rangle=(|r\xi_{+}\rangle\pm|\varpi_{-}\rangle)/\sqrt{2}$. In such a situation, the Hamiltonian $H_{2}^{r}$ in Eq.~(\ref{eq14}) yields
\begin{eqnarray}\label{eq15}
H^{r}_{3}&=&\frac{3\Omega_b^{2}}{8\Delta}(|E_+\rangle+|E_{-}\rangle)(\langle E_+|+\langle E_{-}|)\cr&&
+\Omega_{a}(t)e^{i\varphi_{a}(t)}\big[|\xi_{+}\xi_{-}\rangle\langle r\xi_{-}|+\frac{1}{\sqrt{2}}|\xi_{+}\xi_{+}\rangle(\langle E_+|\cr&&+\langle E_{-}|)+\mathrm{H.c.}\big]
+\frac{\Omega_b}{\sqrt{2}}(|E_+\rangle\langle E_+|-|E_-\rangle\langle E_-|).\cr
&&
\end{eqnarray}
By considering the limit of $\Omega_b\gg\Omega_a(t)$, the collective state  $|\xi_{+}\xi_{+}\rangle$ is decoupled with $|E_{+}\rangle$ and $|E_{-}\rangle$ in the regime of large detuning. It is worth noting that the term $H'_{h2}$ in Eq.~(\ref{eq133}) is very small compared with the Hamiltonian $H_r$ in Eq.~(\ref{eq13}) under the limit condition $\Delta\gg\Omega_{b}\gg\Omega_{a}(t)$, so that we have neglected it in the above derivations. Therefore, the Hamiltonian $H_{3}^{r}$ in 
\mbox{Eq.~(\ref{eq15})} can be further simplified as

\begin{eqnarray}\label{eq16}
H_{\rm eff}=\Omega_{a}(t)e^{i\varphi_{a}(t)}|\xi_{+}\xi_{-}\rangle\langle r\xi_{-}|+\mathrm{H.c.},
\end{eqnarray}
which signifies that the Hamiltonian provides a selective coupling mechanism that only the initial state $|\xi_{+}\xi_{-}\rangle$ can evolve in the subspace $\mathcal{S}$.
 \begin{figure}[htbp] \centering
 	\includegraphics[width=5.5cm]{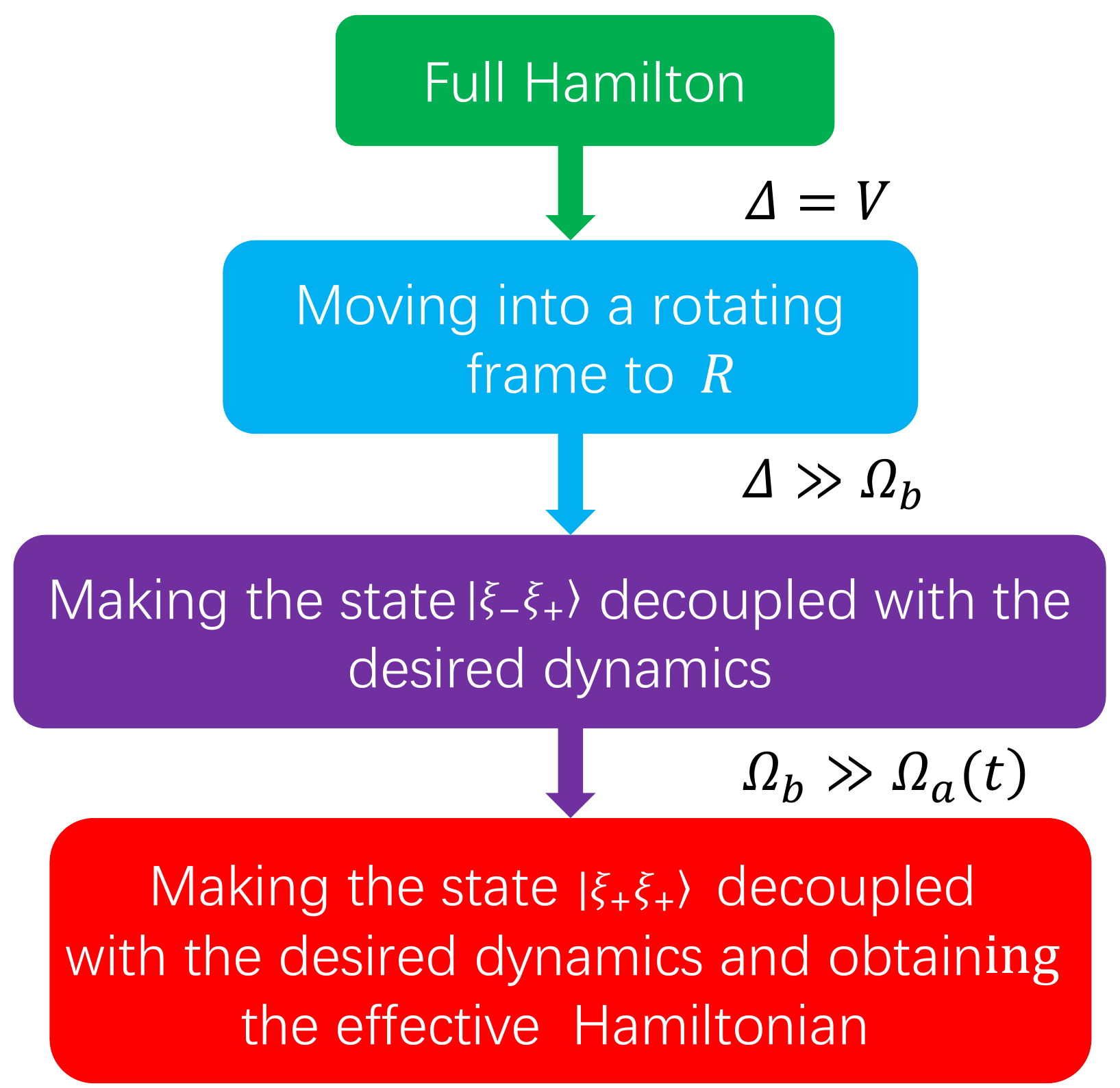}
 	\caption{Sketch map of deriving the effective Hamiltonian.}
 	\label{step}
 \end{figure}
 In a word, such a course of deriving the effective Hamiltonian is shown in Fig. \ref{step}.

\subsection{The realization of NHQC+ gate by using IBRE} \label{III B}
In order to conveniently design the $H_{\rm eff}$ in Eq.~(\ref{eq16}) with IBRE, we denote $\Omega_{x}(t)=2\Omega_{a}(t)\cos\varphi_{a}(t)$, $\Omega_{y}(t)=2\Omega_{a}(t)\varphi_{a}(t)$, and $\Omega_{z}=0$. Therefore, the effective Hamiltonian $H_{\rm eff}$ in Eq. (\ref{eq16}) can be represented as
\begin{eqnarray}\label{eq117}
	H_{\rm eff}=\frac{\Omega_{x}(t)}{2}\sigma_x+\frac{\Omega_{y}(t)}{2}\sigma_y+0\times\sigma_z,
\end{eqnarray}
with $\sigma_x=|r\xi_{-}\rangle\langle \xi_{+}\xi_{-}|+\mathrm{H.c.}$, $\sigma_y=-i|r\xi_{-}\rangle\langle \xi_{+}\xi_{-}|+\mathrm{H.c.}$, and $\sigma_z=|r\xi_{-}\rangle\langle r\xi_{-}|-|\xi_{+}\xi_{-}\rangle\langle \xi_{+}\xi_{-}|$. According to \mbox{Sec.~\ref{II}}, we can make use of IBRE to design the control fields. For a two-level quantum system, the invariant $I(t)$ can be constructed by the superposition of three group generators of SU(2) algebra with time-dependent parameters $\mu_1$ and $\mu_2$  \cite{kang2020heralded},
\begin{eqnarray}\label{eq18}
	I(t)=\frac{u}{2}(\cos\mu_1\sigma_z+\sin\mu_1\sin\mu_2\sigma_x+\sin\mu_1\cos\mu_2\sigma_y),\cr
\end{eqnarray}
where $u$ is an arbitrary constant with units of frequency to keep $I(t)$ with dimensions of energy \cite{PhysRevA.83.062116}. Substituting Eq.~(\ref{eq18}) into Eq.~(\ref{eq1}), $\mu_1(t)$ and $\mu_2(t)$ are satisfied the differential equations as
\begin{eqnarray}\label{eq17}
	\dot{\mu}_1(t)&=&\Omega_y(t)\sin\mu_2(t)-\Omega_x(t)\cos\mu_2(t),\cr\cr
	\dot{\mu}_2(t)&=&\cot\mu_1(t)(\sin\mu_2(t) \Omega_x(t)+\cos\mu_2(t) \Omega_y(t)).\cr
	&&
\end{eqnarray}
By using Eq.~(\ref{eq17}), we can reversely solve the control fields as
\begin{eqnarray}
	\Omega_x(t)&=&\sin\mu_2\tan\mu_1\dot{\mu}_2-\cos\mu_2\dot{\mu}_1
	,\cr\cr
	\Omega_y(t)&=&\cos\mu_2\tan\mu_1\dot{\mu}_2+\sin\mu_2\dot{\mu}_1.
\end{eqnarray}
In addition, the orthogonal eigenvector of the invariant $I(t)$ with the eigenvalues $\pm\mu/2$ can also be obtained as
\begin{eqnarray}\label{eq21}
|\vartheta_1(t)\rangle&=&\cos\frac{\mu_1}{2}|r\xi_{-}\rangle+ie^{-i\mu_2}\sin\frac{\mu_1}{2}|\xi_{+}\xi_{-}\rangle,\cr\cr
|\vartheta_2(t)\rangle&=&ie^{i\mu_2}\sin\frac{\mu_1}{2}|r\xi_{-}\rangle+\cos\frac{\mu_1}{2}|\xi_{+}\xi_{-}\rangle.\cr&&
\end{eqnarray}
According to the results in Eqs.~(\ref{eq10}) and (\ref{eq6}), the time derivatives of dynamic phase and geometric phase acquired for the basis states $|\vartheta_1(t)\rangle$ and  $|\vartheta_2(t)\rangle$ can be calculated as
\begin{eqnarray}\label{22}
	\dot{\theta}_1(t)&=&-\frac{\dot{\mu}_2\sin^2\mu_1}{2\cos\mu_1},\dot{\Theta}_1 (t)=\dot{\mu}_2\sin^2\frac{\mu_1}{2},\cr
	\dot{\theta}_2(t)&=&\frac{\dot{\mu}_2\sin^2\mu_1}{2\cos\mu_1},\dot{\Theta}_2 (t)=-\dot{\mu}_2\sin^2\frac{\mu_1}{2},
\end{eqnarray}
respectively.

To construct NHQC+ dynamics, we choose the auxiliary basis $|\tilde{\vartheta}_1(t)\rangle=|\xi_{-}\xi_{-}\rangle$ and $|\tilde{\vartheta}_2(t)\rangle=|\vartheta_2(t)\rangle$.
Obviously, the state $|\xi_{-}\xi_{-}\rangle$ is satisfied the conditions of NHQC+. Next, we only need to make sure that whether or not the basis $|\vartheta_2(t)\rangle$ satisfies the conditions of NHQC+. First, to satisfy the cyclic condition $|\vartheta_2(0)\rangle=|\vartheta_2(T)\rangle=|\xi_{+}\xi_{-}\rangle$, we consider the boundary conditions satisfy $\mu_1(0)=0$, $\mu_1(T)=0$. In the case, the selection of parameter $\mu_2(t)$ is irrelevant to the cyclic condition according to Eq.~(\ref{eq21}). Second, $|\tilde{\vartheta}_2(t)\rangle\langle \tilde{\vartheta}_2(t)|$ satisfies the von Neumann equation, which is easily proved in Appendix \ref{appb}. In order to make the dynamic phase in the evolution process vanish and obtain a pure geometric phase $\Theta_g$, we design the parameters $\mu_1(t)$ and $\mu_2(t)$ by dividing the time interval $[0,T]$ into two parts including $[0,T/2]$ and $[T/2,T]$. The parameters $\mu_1(t)$ and $\mu_2(t)$ are set as $\mu_1(t)=\mu_1(T-t)$ with $\mu_1(T/2)=\pi$ in the time interval $[0,T]$ and $\mu_2(t)=-\Theta_g+\mu_2(T-t)$ in the time interval $[T/2,T]$, respectively \cite{kang2020heralded}.  With the assumptions of parameters $\mu_1(t)$ and $\mu_2(t)$, the dynamic phase acquired in the evolution time $[0,T]$ can be demonstrated to be equal to zero (see Appendix \ref{appC} for details). 
Thus, the NHQC+ dynamics can be constructed and the evolution operator can be described as
\begin{eqnarray}
	U_0(T,0)=|\xi_{-}\xi_{-}\rangle\langle \xi_{-}\xi_{-}|+e^{i\Theta_g}|\xi_{+}\xi_{-}\rangle\langle\xi_{+}\xi_{-}|,
\end{eqnarray}
i.e., we can achieve the gate operation at $t=T$ when the geometric phase $\Theta_g=\pi$,
\begin{eqnarray}
	U_0(T,0)=
	\begin{bmatrix}
		1 & 0 & 0 & 0\\
		0 & 1 & 0 & 0\\
		0 & 0 & -1 & 0\\
		0 & 0 & 0 & 1\\
	\end{bmatrix},	
\end{eqnarray}
in the subspace $\mathcal{S}$. Furthermore, back the computational basis $\{|00\rangle,|01\rangle,|10\rangle,|11\rangle\}$,  the nontrivial two-qubit NHQC+ gate is represented as
\begin{widetext}
	\begin{eqnarray}\label{eq24}
		U_0(v_a,v_b)=
		\begin{bmatrix}
			\sin^2 v_a+\cos^2 v_a\cos 2v_b & \cos^2 v_a\sin 2v_b & -\sin 2 v_a\sin^2 v_b & \frac{1}{2}\sin 2v_a\sin 2v_b\\\\
			\cos^2 v_a\sin 2v_b & \sin^2 v_a-\cos^2 v_a\cos 2v_b & \frac{1}{2}\sin 2v_a\sin 2v_b & -\sin 2v_a\cos^2 v_b\\\\
			-\sin 2 v_a\sin^2 v_b & \frac{1}{2}\sin 2v_a\sin 2v_b & \cos^2 v_a+\sin^2 v_a\cos 2v_b & \sin^2 v_a\sin 2v_b\\\\
			\frac{1}{2}\sin 2v_a\sin 2v_b &  -\sin 2v_a\cos^2 v_b & \sin^2 v_a\sin 2v_b & \cos^2 v_a-\sin^2 v_a\cos 2v_b\\\\
		\end{bmatrix}.\cr
	&&	
	\end{eqnarray}
\end{widetext}
	
By choosing different values of $v_a$ and $v_b$, different two-qubit NHQC+ gates
$U_0(v_a,v_b)$ in Eq.~(\ref{eq24}) are realized.	
For $v_a=\pi/2$, $v_b=\pi$, $U_0(\pi/2,\pi)$ is

\begin{eqnarray}
	U_0(\pi/2,\pi)=
	\begin{bmatrix}
		1 & 0 & 0 & 0\\
		0 & 1 & 0 & 0\\
		0 & 0 & 1 & 0\\
		0 & 0 & 0 & -1\\	
	\end{bmatrix},	
\end{eqnarray}
which is a holonomic controlled phase. While for $v_a=\pi/2$, $v_b=\pi/4$, a holonomic controlled-NOT (CNOT) gate
$U_0(\pi/2,\pi/4)$ is

\begin{eqnarray}
	U_0(\pi/2,\pi/4)=
	\begin{bmatrix}
		1 & 0 & 0 & 0\\
		0 & 1 & 0 & 0\\
		0 & 0 & 0 & 1\\
		0 & 0 & 1 & 0\\	
	\end{bmatrix}.	
\end{eqnarray}

Until now, we have constructed the NHQC+ dynamics and implemented nontrivial two-qubit NHQC+ gate. However, the systematic errors of control fields always exist in the experiment, which will inevitably reduce the fidelity of gate.
The previous schemes of the NHQC are sensitive to the systematic errors and difficult to incorporate optimal control technique without additional adjustable parameters. Here, for improving the fidelity of the NHQC+ gate, we introduce zero-systematic-error-sensitivity optimal control method to further enhance the robustness of the systematic error in the implementation of quantum gate \cite{Ruschhaupt2012}.
In the presence of the systematic error $\varepsilon$ of the control fields,
the effective Hamiltonian in Eq. (\ref{eq117}) can be written as
\begin{eqnarray}
	H_{\rm eff}=(1+\varepsilon)\frac{(\Omega_x+i\Omega_y)}{2}|\xi_{+}\xi_{-}\rangle\langle r\xi_{-}|+\mathrm{H.c.}
\end{eqnarray}
By using perturbation theory up to $O(\varepsilon^2)$, we can obtain
$|\psi^{\varepsilon}_{2}(T)\rangle=|\psi_2(T)\rangle-i\varepsilon\int_{0}^{T}dtU_0(T,t)H_{\rm eff}(t)|\psi_2(t)\rangle
-\varepsilon^2\int_{0}^{T}dt\int_{0}^{t}dt^{\prime}U_0(T,t)H_{\rm eff}(t)U_0(t,t^{\prime})H_{\rm eff}(t^{\prime})|\psi_2(t^{\prime})\rangle+O(\varepsilon^3)$ \cite{sjoqvist2012non}, where $|\psi_2(t)\rangle$ ($|\psi^{\varepsilon}_{2}(t)\rangle$) are the state of the system in the absence (presence) of systematic errors.  
As the evolution of system can be described by $|\psi_{2}(t)\rangle=e^{i\alpha_2(t)}|\vartheta_2(t)\rangle$, the fidelity of evolution is estimated as
\begin{eqnarray}
	P_{\varepsilon}&=&\left|\langle \psi_2(T)|\psi_2^{\varepsilon}(T)\rangle\right|^2\cr
	&=&1-\varepsilon^2\left|\int_{0}^{T}dte^{2i\alpha_2(t)}\langle \vartheta_1(t)|H_{\rm eff}(t)|\vartheta_2(t)\rangle\right|^2,\cr
	&&
\end{eqnarray}
with $\alpha_1(t)=-\alpha_2(t)$ being considered. 

To evaluate the influence of the static systematic error, the systematic error sensitivity \cite{Ruschhaupt2012}  defined as
$q_s=-\frac{\partial P_{\varepsilon}}{2\partial \varepsilon^2}$ is introduced. The systematic error sensitivity can be calculated as
\begin{eqnarray}
	q_s=-\frac{\partial P_{\varepsilon}}{2\partial \varepsilon^2}=|\int_{0}^{T}dt\exp(i\chi)\dot{\mu}_1\sin^2\mu_1|^2,
\end{eqnarray}
with $\chi(t)=\mu_2(t)+2\alpha_2(t)$. Here, we set $\chi(\mu_1)=
\eta[2\mu_1-\sin(2\mu_1)]$ to minimize the $q_s$ with $\eta$ being a time-independent optimized coefficient. In order to make the improvement of the gate performance attributed to the optimal control, the maximum value of the optimized control fields is bounded by the maximal amplitude $\Omega_{max}$ of control fields $\Omega_{x}(t)$ and $\Omega_{y}(t)$.
Under the restriction of $\Omega_{max}$, the evolution time $T$ increases with the value of $\eta$, which aggravates the influence of decoherence on the fidelity of the quantum gate. For this choice of $\chi$, we can obtain $q_s=\frac{\sin^2(\eta \pi)}{\eta^2}$. Clearly, when $\eta$=1, 2, 3, ..., the value of $q_s$ equal to zero, which means that the minimum of the systematic error sensitivity is achieved. To trade off between the
robustness against systematic errors and decoherence, we choose the optimized coefficient $\eta=1$. Accordingly, we have $\mu_2(t)=4\sin^3\mu_1(t)/3$ in the time interval $[0,T/2]$ and $\mu_2(t)=4\sin^3\mu_1(t)/3-\pi$ in the time interval $[T/2,T]$. To make the control fields continuous and vanish at the boundary, we choose $\mu_1(t)=\pi\sin^2(\pi t/T)$ in the time interval $[0,T]$.
\begin{figure}[hbpt]
	\centering
	\includegraphics[width=1\linewidth]{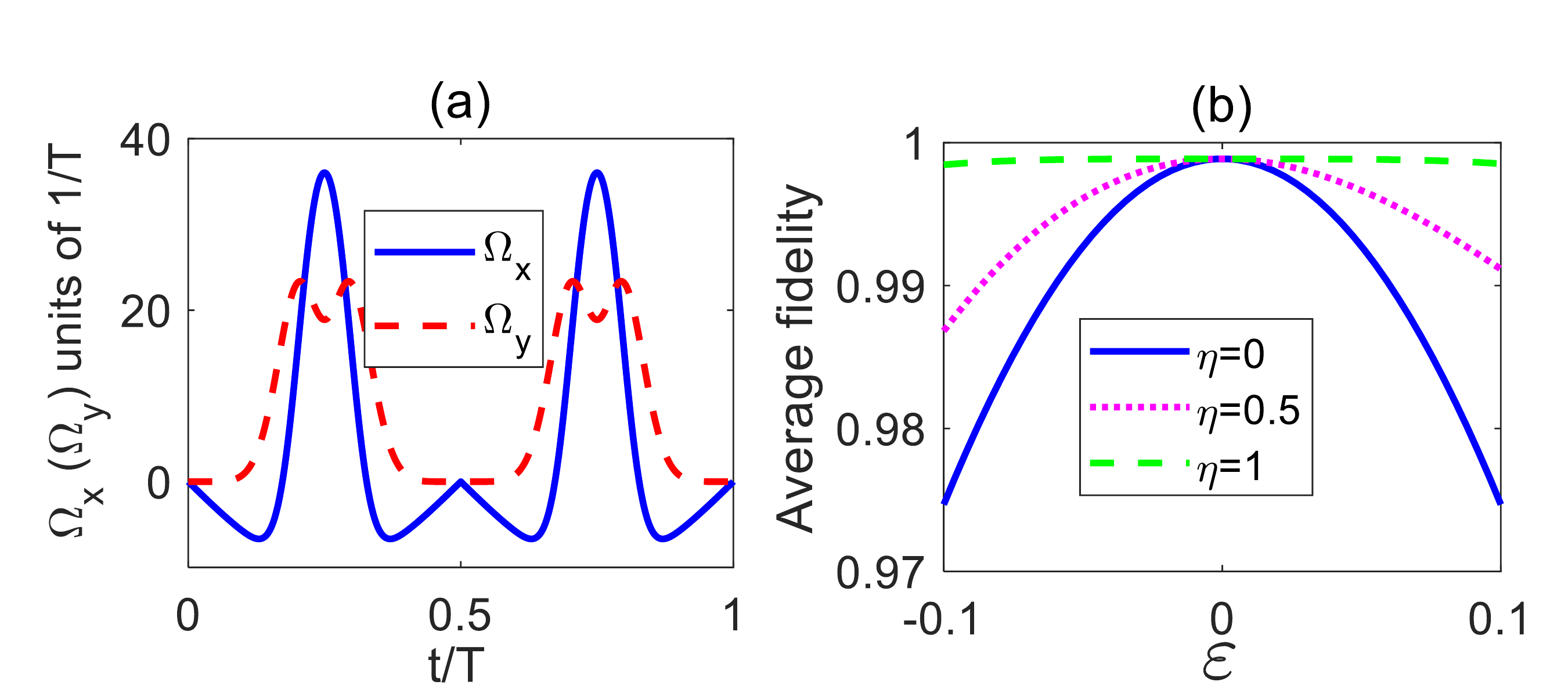}
	\caption{(a) Dependences on $t$ of control fields $\Omega_{x}(t)$, $\Omega_{y}(t)$ with $\eta=1$. (b) The performance of the average fidelity of the holonomic CNOT gate with systematic error $\varepsilon$ for different optimized coefficients $\eta$.}
	\label{fig3}
\end{figure}

\begin{figure}[hbpt]
 	\centering
 	\includegraphics[width=1\linewidth]{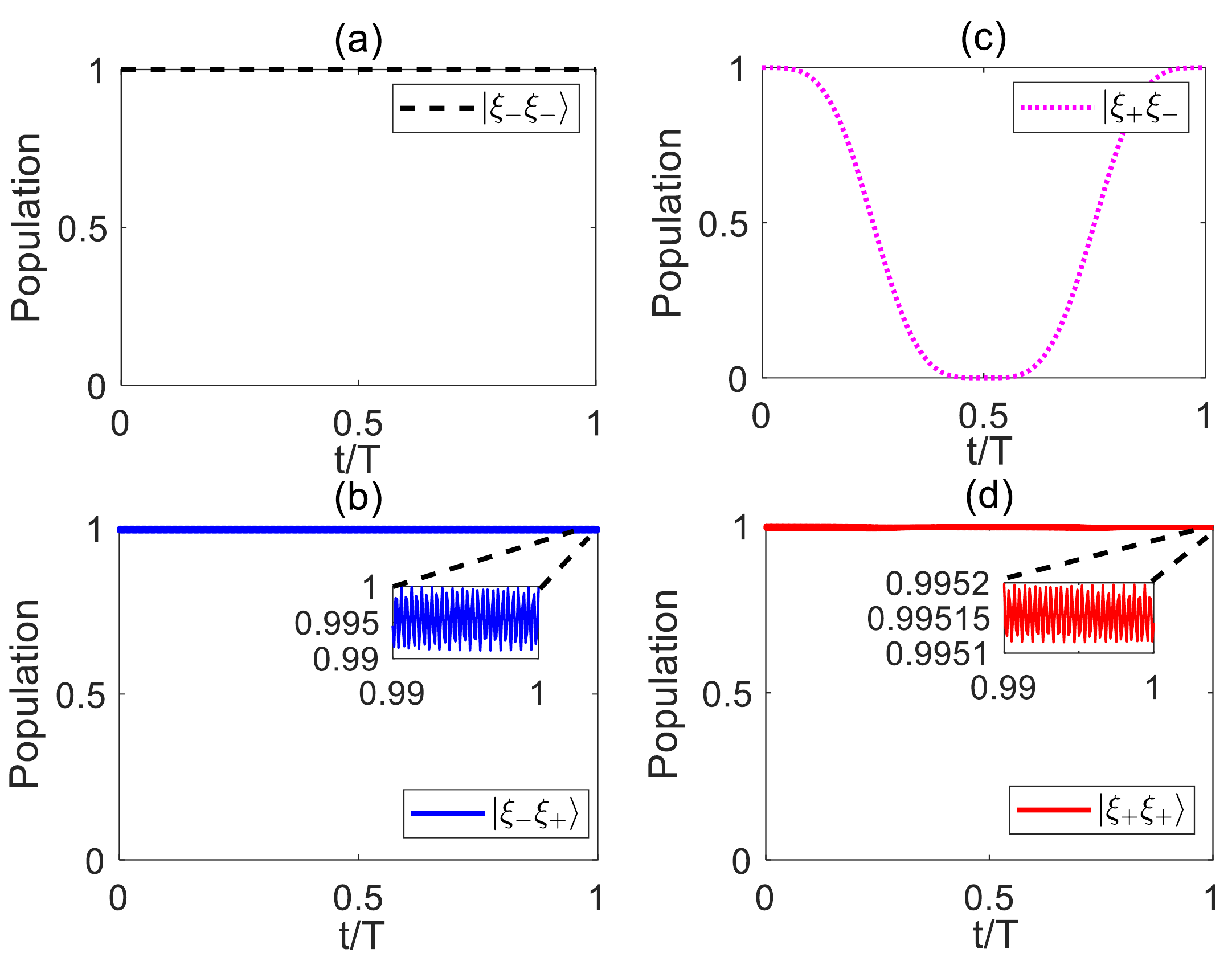}
 	\caption{The time evolution of populations for the states $|\xi_{-}\xi_{-}\rangle$, $|\xi_{-}\xi_{+}\rangle$, $|\xi_{+}\xi_{-}\rangle$ and $|\xi_{+}\xi_{+}\rangle$ in the subspace $\mathcal{S}$ in (a)-(d), respectively. The time evolution of population is defined as $Q^{\prime}=|\langle\Phi_{0}^{\prime}(T)|\Phi_0^{\prime}(t)\rangle|^{2}$ for the states in the subspace $\mathcal{S}$, where $|\Phi_{0}^{\prime}(T)\rangle$ is the target state and $|\Phi_0^{\prime}(t)\rangle$ is the state of system at time $t$. The parameters are set $\eta=1$, $V=2\pi\times133.04$ MHz, $\Omega_b=2\pi\times4.43$ MHz, and $\Omega^{max}_{0}=2\pi\times0.27 $ MHz.}
 	\label{figff}
 \end{figure}

With the choice of the parameters designed above, we plot the control fields $\Omega_{x,y}(t)$ versus $t$ in Fig. \ref{fig3}(a). From Fig.~\ref{fig3}(a), we can further obtain the maximal value of control field $\Omega_{max}=36.05/T$.
 In order to better satisfy the condition $V\gg\Omega_{b}$, the Rydberg interaction strength is set as $V=18000/T$ and the control field $\Omega_b=600/T$ to build up the effective Hamiltonian. In this case, the total operation time is $T=18000/V$. For the dipole-dipole interaction $V$, its strength $\sqrt{2}C_3/R^3$ can be continuously varied in the range of $2\pi\times [3.38, 216.32]$ MHz \cite{Ravets2014,PhysRevA.92.020701}, where $C_3=2\pi\times2.39$ GHz $\mu {\rm m}^3$ and $R$ represents the interatomic distance. Besides, the Rabi coupling strength of the ground state and the Rydberg state can be up to $2\pi\times5$ MHz realized by a two-photon process \cite{PhysRevA.92.020701}.  Here, we consider the experimentally feasible parameter $V=2\pi\times133.04$ MHz at the distance of $R\approx2.94$ $\mu$m, leading to short evolution time $T=21.5$ $\mu$s. Correspondingly, the control field $\Omega_{b}=2\pi\times4.43$ MHz and the maximal value of control field $\Omega^{max}_{0}=2\pi\times0.27$ MHz, which is available in the experiment. Under the current parameter condition, it is important to verify the validity of the effective Hamiltonian $H_{\rm eff}$ in  Eq.~(\ref{eq117}). In Figs.~\ref{figff}(a)-\ref{figff}(d), we plot the time evolution of populations for the states $|\xi_{-}\xi_{-}\rangle$, $|\xi_{-}\xi_{+}\rangle$, $|\xi_{+}\xi_{-}\rangle$ and $|\xi_{+}\xi_{+}\rangle$ in the subspace $\mathcal{S}$ governed by the full Hamiltonian in Eq.~(\ref{eq9}), respectively.
 It can be seen that this result is in good agreement with the expectation. As shown in Fig.~\ref{figff}, the states $|\xi_{-}\xi_{-}\rangle$, $|\xi_{-}\xi_{+}\rangle$, and $|\xi_{+}\xi_{+}\rangle$ always keep in their initial states with higher fidelities and the state $|\xi_{+}\xi_{+}\rangle$ can realize the coherent population transfer. Therefore, the validity of effective Hamiltonian $H_{\rm eff}$ in Eq.~(\ref{eq117}) is verified.

  \begin{figure}[hbpt]
 	\centering
 	\includegraphics[width=1\linewidth]{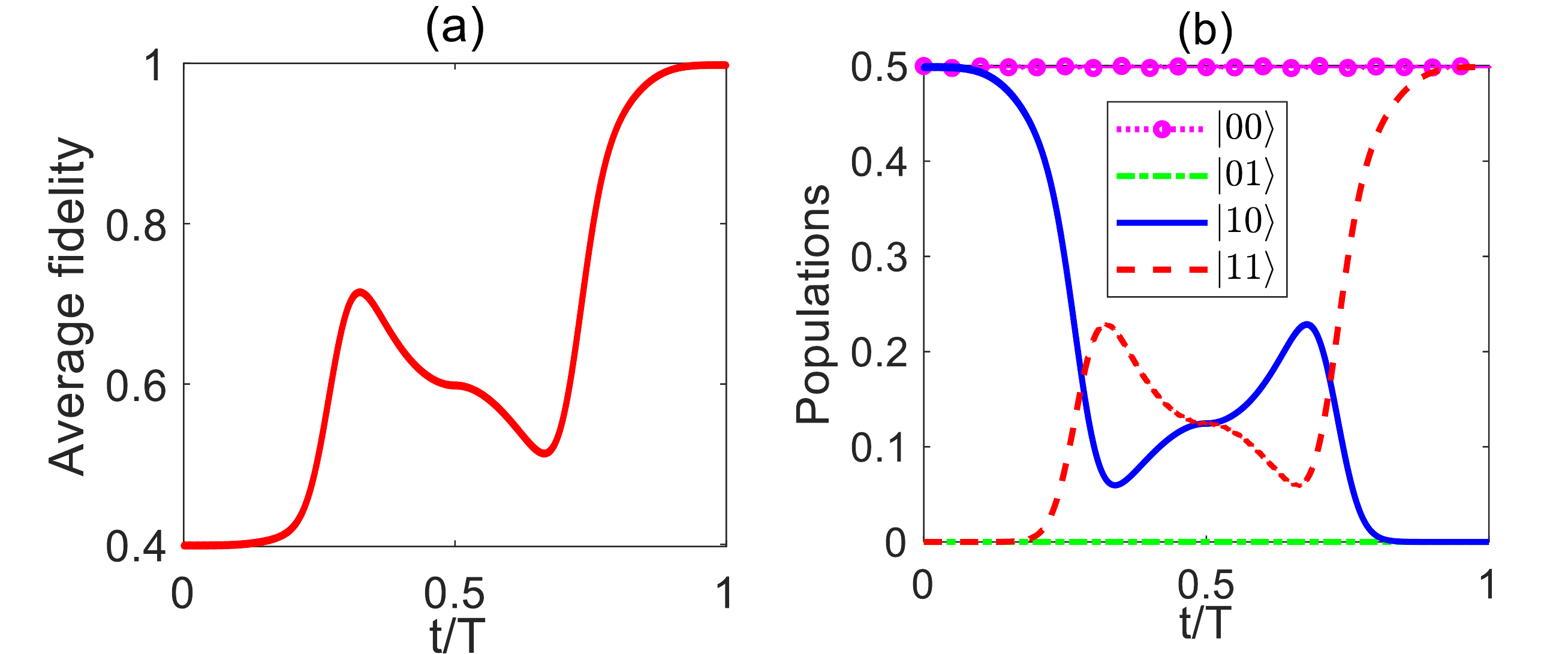}
 	\caption{(a) Average fidelity $F_{\rm CN}$ of the implementation of the holonomic CNOT gate versus $t$ with the full Hamiltonian in Eq.~(\ref{eq9}), where we choose $\eta=1$. (b) The time evolution of population for the computational basis states $|00\rangle$, $|01\rangle$, $|10\rangle$ and $|11\rangle$. The time evolution of population is defined as $Q=|\langle\Phi_{0}(T)|\Phi_0(t)\rangle|^{2}$ for the computational basis states, where $|\Phi_{0}(T)\rangle$ is the target state and $|\Phi_0(t)\rangle$ is the state of system at time $t$. Corresponding parameters are the same as that in Fig.~\ref{figff}.}
 	\label{fig4}
 \end{figure}

As an example, we check the robustness of the implementation of the holonomic CNOT gate against systematic errors of control fields and show the results of optimization by numerical simulation.
The final average fidelities $F_{\rm CN}(T)$ versus error $\varepsilon$ with different optimized coefficient $\eta$ are plotted in Fig.~\ref{fig3}(b), where $F_{\rm CN}(t)$ is defined as \cite{PhysRevA.70.012315,Pedersen2007}
\begin{eqnarray}
	F_{\rm CN}(t)=\frac{1}{N(N+1)}\{Tr[M(t)M(t)^{\dagger}]+|Tr[M(t)]|^2\},\cr
\end{eqnarray}
with $M(t)=P_cU_{\rm CN}^{\dagger}U(t)P_c$, $U_{\rm CN}=|00\rangle\langle 00|+|01\rangle\langle01|+|10\rangle\langle11|+|11\rangle\langle10|$,
$P_{c}=|00\rangle\langle 00|+|01\rangle\langle01|+|10\rangle\langle10|+|11\rangle\langle11|$ being the projection operator onto the computational subspace, and $N=4$ representing the dimension of the computational subspace. Seen from the blue solid line in Fig.~\ref{fig3}(b), $F_{\rm CN}(T)$ with the optimized coefficient $\eta=1$ is always higher than 99.8$\%$ when $\varepsilon\in[-0.1,0.1]$,
which shows that the implementation of the holonomic CNOT gate is quite insensitive to systematic errors. We also see that the average fidelity falls to 0.9747 when $\varepsilon=\pm0.1$ with $\eta=0$. Noticing that when $\eta=0$ the current scheme reduces to previous NHQC cases \cite{PhysRevA.92.022320,PhysRevA.94.022331,PhysRevApplied.7.054022,PhysRevA.96.052316,PhysRevLett.121.110501}. Thus, the scheme can enhance the robustness against systematic errors.
 \begin{figure}[hbpt]
	\centering
	\includegraphics[width=0.9\linewidth]{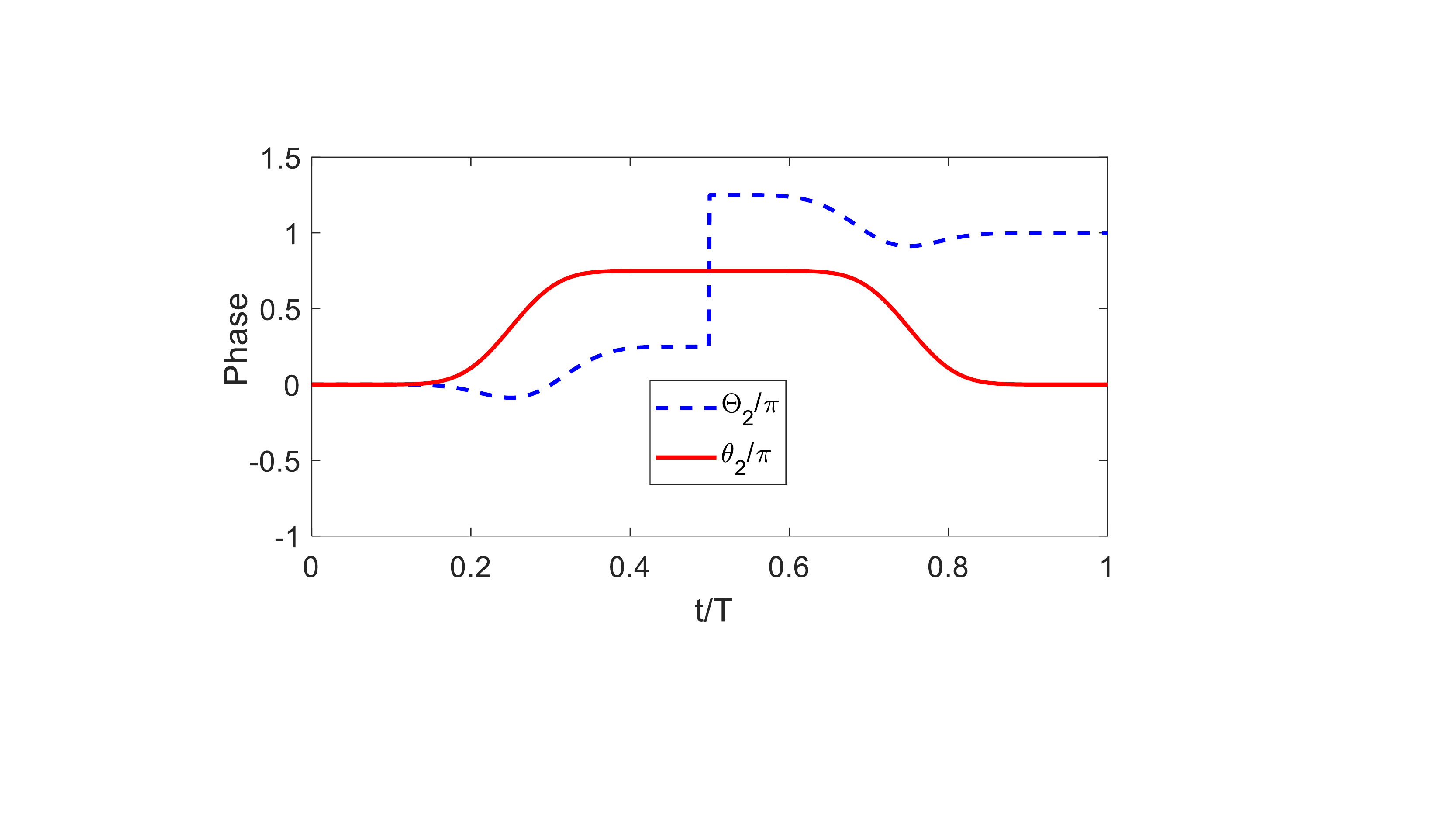}
	\caption{The geometric phase $\Theta_2(t)$ and the dynamic phase $\theta_2(t)$ acquired in the implementation of the holonomic CNOT gate versus $t$ with the effective Hamiltonian $H_{\rm eff}$ in Eq. (\ref{eq117}). Corresponding parameters are the same as that in Fig.~\ref{figff}.}
	\label{phase}
\end{figure}
In addition, we also simulate the average fidelity $F_{\rm CN}(t)$ in Fig.~\ref{fig4}(a). As seen from Fig.~\ref{fig4}(a), we obtain the average fidelity of the holonomic CNOT gate as $F_{\rm CN}(T)=99.89\%$ at $t=T$, which is in accordance with the expectation. Specially, we also choose a special initial state $|\Phi_0(0)\rangle=\frac{1}{\sqrt{2}}(|00\rangle+|11\rangle)$ to verify the validity of the CNOT gate. Correspondingly, the target state is $|\Phi_0(T)\rangle=\frac{1}{\sqrt{2}}(|00\rangle+|11\rangle)$. As shown in Fig.~\ref{fig4}(b), the computational basis states $|00\rangle$ and $|01\rangle$ keep in the initial states and the states $|10\rangle$ and $|11\rangle$ accomplish the coherent population transfer. Moreover, the acquired geometric phase $\Theta_2(t)$ and dynamic phase $\theta_2(t)$ are plotted in Fig.~\ref{phase}. We can see from Fig.~\ref{phase} that the dynamic phase $\theta_2(T)$ finally vanishes at $t=T$, while the geometric phase reaches predetermined value
 $\Theta_2(T)=\Theta_g=\pi$, which means that  a pure geometric phase is acquired in the evolution process. Therefore, the holonomic CNOT gate can be successfully realized with high fidelity.

\section{Analysis of errors and decoherence on the fidelity of the NHQC+ gate}\label{IV}
In addition to the systematic errors $\varepsilon$ of the control fields discussed in Sec.~\ref{III B},
there may exists other errors and decoherence affecting the performance of the NHQC+ gate. In a practical experiment environment, the analysis of the robustness against these disturbing factors is helpful for estimating the performance of the scheme.
As an example, we analyze the robustness for implementation of the holonomic CNOT gate in the following discussion.
\begin{figure*}[hbpt]
	\centering
	\includegraphics[width=0.8\linewidth]{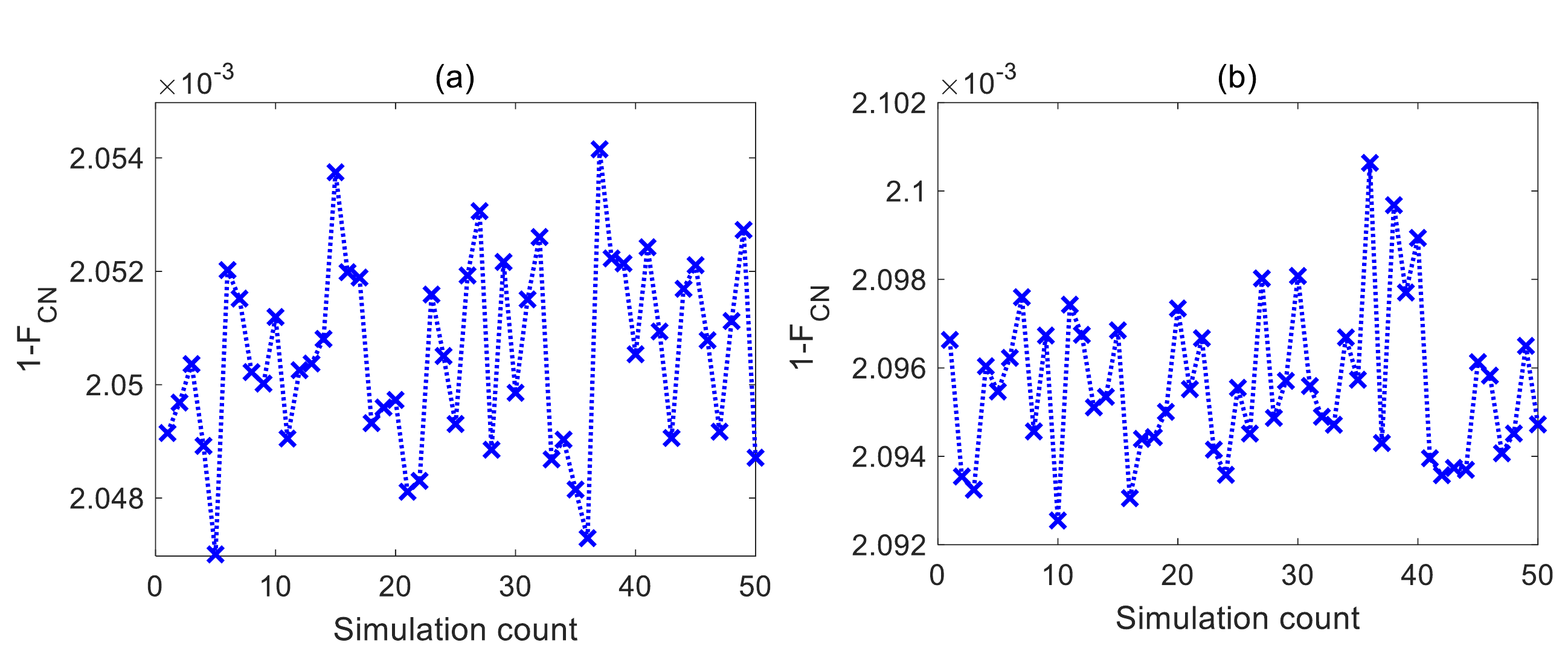}
	\caption{$1-F_{\rm CN}$ versus simulation count with signal-to-noise ratios (a) $SNR=10$ and (b) $SNR=2$, respectively. Corresponding parameters are the same as that in Fig.~\ref{figff}.}
	\label{figa}
\end{figure*}
 \subsection{Random noise in control fields}
  In the experiment, the random noise is an error source reducing the average fidelity of gate in experiment. Because the fluctuation of Rabi frequencies of the control fields is always random, thus it is worthwhile to study the implementation of gate in a noisy environment. As the additive white Gaussian noise (AWGN) is a nice model to simulate the process of many random noises, here we study the performance of gate by adding AWGN to the original control fields. The control fields under the influence of the AWGN can be described as
\begin{eqnarray}
 	\Omega^{\prime}_{x}(t)&=&\Omega_{x}(t)+{\rm AWGN}(\Omega_x(t),SNR),\cr\cr
 	\Omega^{\prime}_{y}(t)&=&\Omega_{y}(t)+{\rm AWGN}(\Omega_y(t),SNR),
\end{eqnarray}
where ${\rm AWGN}(\Omega_{x}(t),SNR)$ (${\rm AWGN}(\Omega_{y}(t),SNR))$ denotes a function generating AWGN with the signal-to-noise ratio $R$ for control field $\Omega_{x}(t)$ ($\Omega_{y}(t)$). Because the AWGN is random in every single simulation, we perform 50 numerical simulations for the implementation of the CNOT gate with $SNR=10$ and $SNR=2$, respectively. The infidelities $1-F_{\rm CN}$ for corresponding signal-to-noise in each simulation are shown in Figs.~\ref{figa}(a) and \ref{figa}(b), respectively. As seen from Fig.~\ref{figa}, the infidelities $1-F_{\rm CN}$ fluctuate around 0.00205 and 0.00209 with $SNR=10$ and $SNR=2$, respectively. Compared with the case without AWGN, the noise only reduces the fidelity of the gate by $10^{-3}$. Therefore, the results show that the scheme is robust against AWGN. The physical mechanism behind this phenomenon is that the random values of AWGN added to the original control fields are zero average, thus the total effect of AWGN on the evolution of system can be nearly neglected. As the accumulations of geometric phases are mainly related to the global property of the evolution path, the local fluctuations caused by AWGN can not cause more errors \cite{PhysRevA.101.032322}.
\begin{figure}[hbpt]
	\centering
	\includegraphics[width=0.9\linewidth]{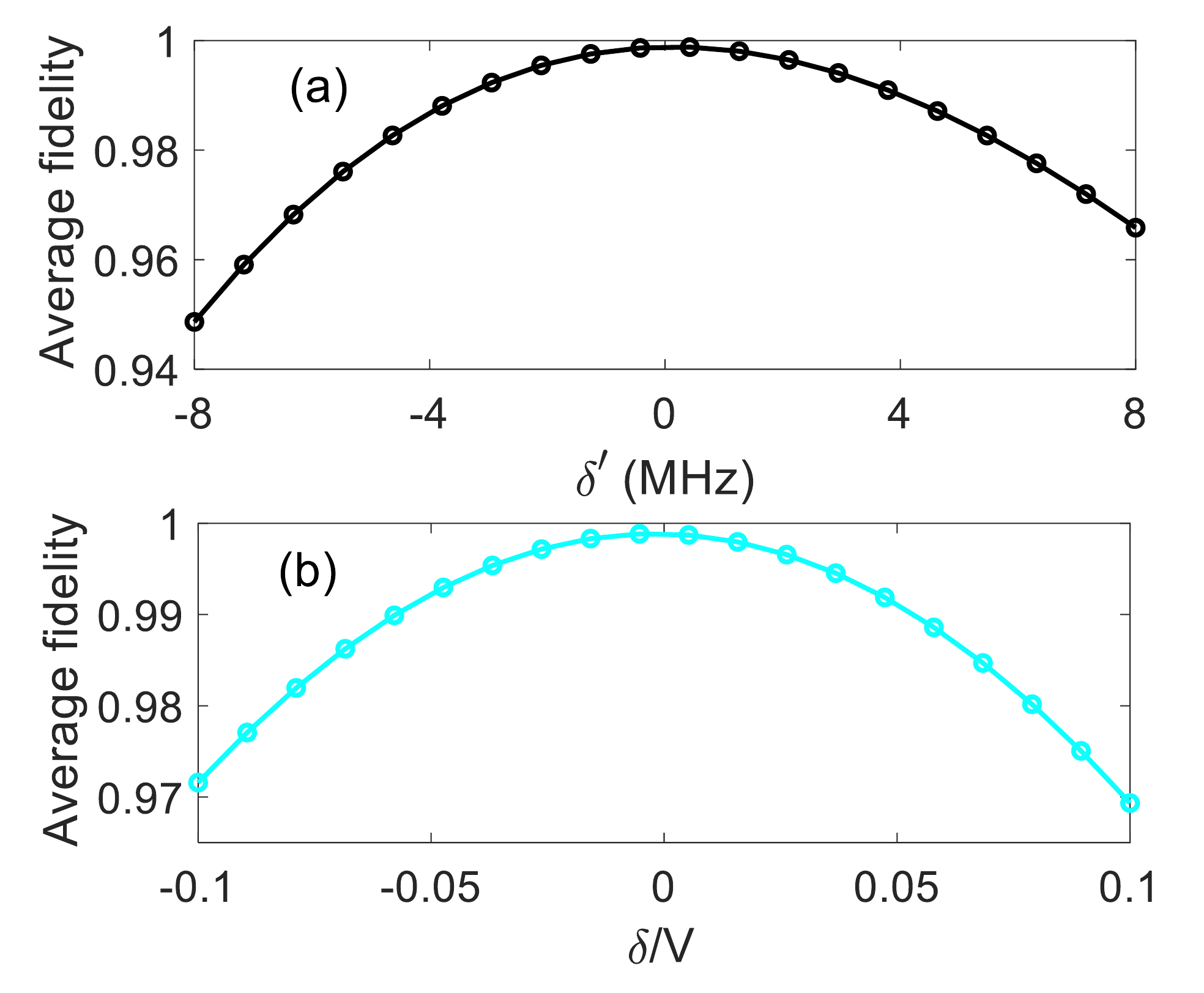}
	\caption{(a) The effect of deviation $\delta^{\prime}$ on the average fidelity of the holonomic CNOT gate. (b) The effect of the F\"orster defect $\delta$ on the average fidelity of the holonomic CNOT gate. Corresponding parameters are the same as that in Fig.~\ref{figff}.}
	\label{fig7}
\end{figure}
\subsection{Deviation of dipole-dipole interaction}
In the process of obtaining the effective Hamiltonian $H_{\rm eff}$ in Eq.~(\ref{eq16}), we have assumed the condition $\Delta=V$. Nevertheless, it is difficult to strictly control the interaction between Rydberg states to satisfy the condition in the experimental operation. In order to evaluate the performance of deviation of dipole-dipole interaction on the average fidelity of the holonomic CNOT gate, we plot the average fidelity of the gate versus the deviation $\delta^{\prime}$ in Fig.~\ref{fig7}(a), where we suppose $V=2\pi\times(133.04+\delta^{\prime})$ MHz. As can be seen from Fig.~\ref{fig7}(a), the current scheme is insensitive to the movement of the distance between atoms, and the average fidelity of the gate is higher than 94.5$\%$. This infidelity is mainly due to the term $\frac{\Omega_b}{\sqrt{2}}|r\xi_{+}\rangle\langle \varpi_{-}|+\mathrm{H.c.}$ of $H_r$ in Eq.~(\ref{eq13}). When the deviation of dipole-dipole interaction exists, the transition between $|r\xi_{+}\rangle$ and  $|\varpi_{-}\rangle$ becomes not resonant. Under the condition $\Omega_b\gg\Omega_a(t)$, the state $|\xi_{+}\xi_{+}\rangle$ can not be completely suppressed in the evolution process.
\subsection{F\"orster defect}
In the presence of F\"orster defect, the interaction Hamiltonian between atoms $H_F$ in Eq.~(\ref{eq99}) is modified in the basis $\{|rr\rangle,|R\rangle\}$ by
\begin{eqnarray}
H'_{F}=
	\begin{pmatrix}
		0 & V \\
		V & \delta \\	
	\end{pmatrix},	
\end{eqnarray}
where $\delta$ is the F\"orster defect between states $|rr\rangle$ and $|R\rangle$. The F\"orster defect changes the eigenvalue with a small deviation $|V-\frac{1}{2}(\delta+\sqrt{4V^2+\delta ^2})|$, which is approximately equal to $\delta/2+\delta ^2/8V$ for a small ratio $\delta/V$. In Fig. \ref{fig7}(b), we numerically simulate the effect of the F\"orster defect on the average fidelity of the holonomic CNOT gate. As shown in Fig.~\ref{fig7}(b), the average fidelity still reaches about 97$\%$ when $\delta/V=\pm0.1$. Considering the experimental parameter of defects $\delta=2\pi\times8.5$ MHz in the absence of an electric field \cite{Ravets2014}, the average fidelity of the gate is 98.64$\%$, which shows that the scheme is still effective.
On the whole, the scheme is robust against the deviation of F\"orster defect.

\subsection{Decoherence induced by dissipation}
\begin{figure}[hbpt]
	\centering
	\includegraphics[width=1\linewidth]{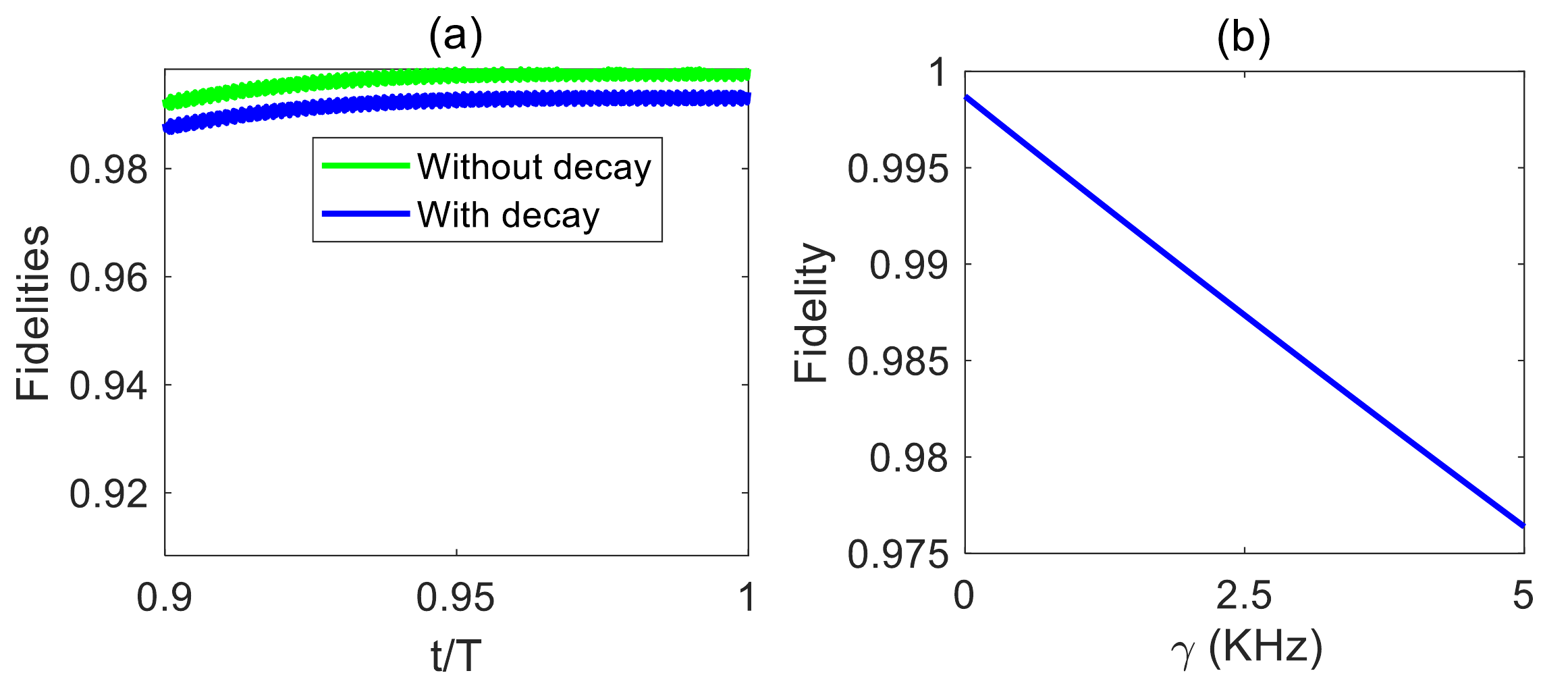}
	\caption{(a) Fidelities $F_{\rm CN}$ of the implementation of the holonomic CNOT gate versus $t$ with decay $\gamma=1$ kHz and without decay by choosing initial state $(|00\rangle+|10\rangle)/\sqrt{2}$. (b) Final fidelities $F_{\rm CN}(T)$ of the holonomic CNOT gate versus decay rate $\gamma$ with initial state $(|00\rangle+|10\rangle)/\sqrt{2}$. Corresponding parameters are the same as that in Fig. \ref{figff}.}
	\label{fig5}
\end{figure}
So far we do not consider the influence of decoherence
on the scheme.
In the practical implementation, the system is always inevitably affected by the environment. Therefore,
we check the performance of decoherence on the implementation of the holonomic CNOT gate with the initial state $(|00\rangle+|10\rangle)/\sqrt{2}$ in the end of this section. 	
The dissipation dynamics process of the system is govern by the master equation,
\begin{eqnarray}
	\dot{\rho}(t)&=&i\big[H_{I},\rho(t)\big]+\sum_{p=1}^{2}\sum_{p'=r, r_{+}, r_{-}}\big[L_{pp'}\rho(t)L_{pp'}^{\dagger}-\cr
	&&-\frac{1}{2}L^{\dagger}_{pp'}L_{pp'}\rho(t)-\frac{1}{2}\rho(t)L^{\dagger}_{pp'}L_{pp'}\big],	
\end{eqnarray}
where $\rho(t)$ being the density operator of the system, $p$ being the atomic number, and $p'$ being the Rydberg state. Besides, the corresponding decay from excited states $|r\rangle$, $|r_{+}\rangle$, $|r_{-}\rangle$ into ground states $|0\rangle$ and $|1\rangle$
 can be expressed as the operator $L_{pp'}=\sqrt{\gamma_{pp^{\prime}}}|j\rangle_{p}\langle p'|$ $(j=0,1$). For simplicity, we assume the atomic spontaneous emission rates $\gamma_{pp^{\prime}}=\gamma$.
\begin{figure}[hbpt]
	\centering
	\includegraphics[width=0.9\linewidth]{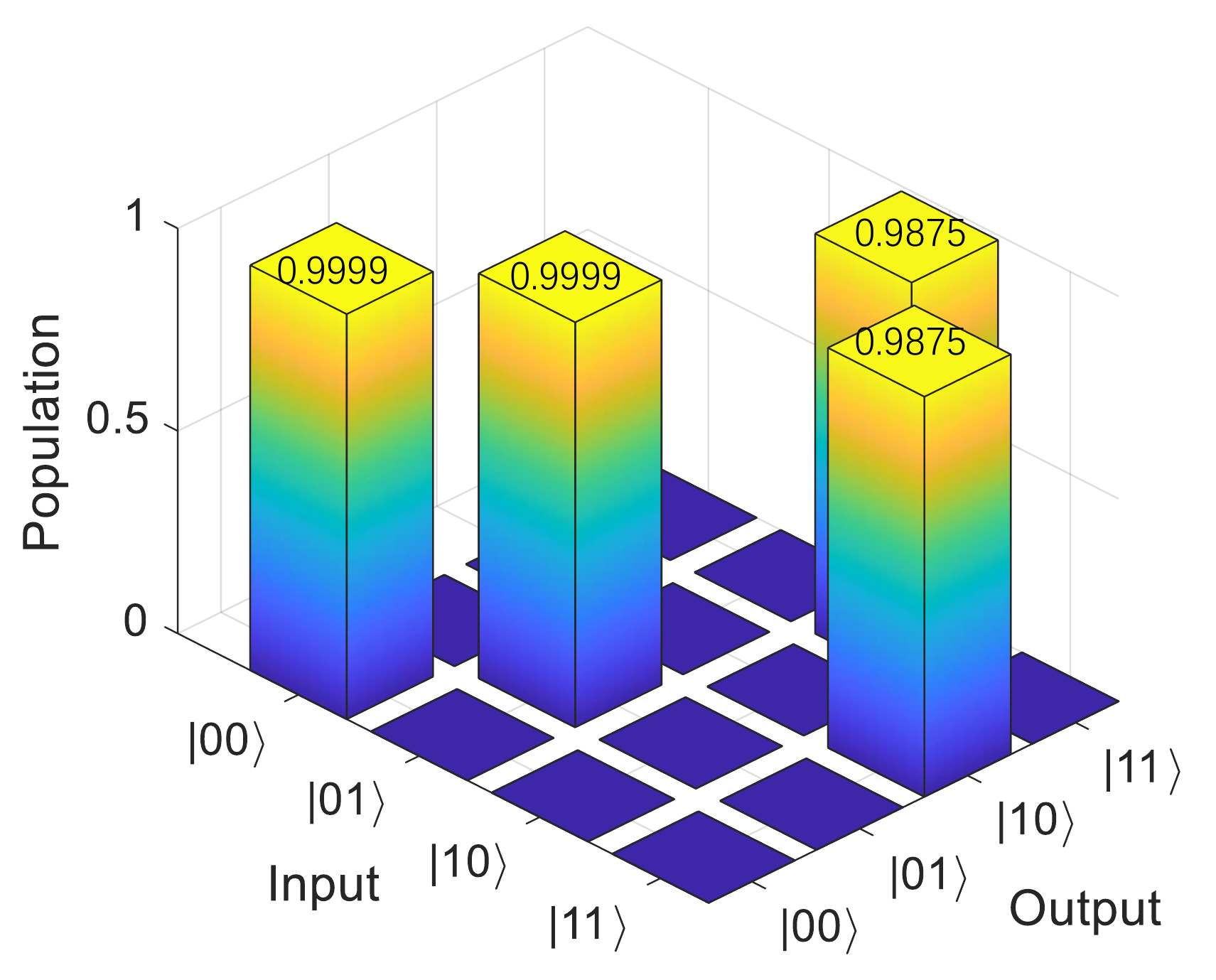}
	\caption{Truth table for the scheme of the holonomic CNOT gate in the presence of decay $\gamma=1$ kHz. Corresponding parameters are the same as that in Fig. \ref{figff}.}
	\label{fig6}
\end{figure}
We plot the fidelities of the evolution of initial state versus $t$ without decay and with decay in Fig.~\ref{fig5}(a). As seen from the green solid line in Fig.~\ref{fig5}(a), the fidelity of the holonomic CNOT gate is $F_{\rm CN}(T)=99.89\%$ at $t=T$ without decay. Moreover, the blue solid line in Fig.~\ref{fig5}(a) represents the fidelity versus $t$ in the presence of decay with $\gamma=1$  kHz, which indicates $F_{\rm CN}(T)=99.42\%$. We also plot the final fidelity $F_{\rm CN}(T)$ of the holonomic CNOT gate with the initial state
$(|00\rangle+|10\rangle)/\sqrt{2}$ versus decay $\gamma$ in Fig.~\ref{fig5}(b). From Fig.~\ref{fig5}(b), we can find the final fidelity $F_{\rm CN}(T)$ is very high (above $98\%$). Thus, the scheme is robust against the atomic spontaneous emission. This is because the populations of partial Rydberg states are decoupled with the effective Hamiltonian in the limits of large detuning $\Delta\gg\Omega_{b}$ and $\Omega_{b}\gg\Omega_{a}(t)$.
In order to better characterize the performance of the holonomic CNOT gate in the presence of decay, we also numerically calculate the truth table \cite{reed2012realization} in Fig.~\ref{fig6}. Specifically, we need to perform the gate operation on each of the computational basis states (Input state $|\Phi_{In}\rangle$). By calculating the populations defined as $|\langle \Phi_{Out}|U_{\rm CN}(T)|\Phi_{In}\rangle|^2$, we can obtain the corresponding population of output state ($|\Phi_{Out}\rangle$) for each input state, where $U_{\rm CN}(T)$ is the evolution operator for the gate at $t=T$. From the numbers on the bars shown in Fig.~\ref{fig6}, the minimum values of population of each input state exceeds 98.7$\%$. Thus, the scheme is helpful to realize a perfect two-qubit gate operation in Rydberg atom systems.

\section{Conclusion}\label{VI}
In conclusion, we have proposed a scheme to realize the NHQC+ in the regime of Rydberg F\"orster resonance. In this scheme, assisted by the strong dipole-dipole interaction between Rydberg atoms, a selective coupling mechanism for the subspace $\mathcal{S}$ is proposed to simply the dynamics of system. In the whole evolution process, the simultaneous excitation of Rydberg atoms is avoided. Based on the derived effective Hamiltonian, the NHQC+ dynamics can be naturally constructed and the corresponding control fields are designed to implement the quantum gate with the help of IBRE. The advantages
of IBRE in NHQC+ have been
shown in the paper. On one hand, the eigenvectors of the
dynamical invariant provide natural evolution paths to realize NHQC+.
On the other hand, IBRE can be
compatible with the systematic-error-sensitivity optimal control
method, which further enhances the robustness against systematic
errors of control fields. By minimizing the systematic error sensitivity $q_s$,
the optimal Rabi frequencies of control fields can be designed. Moreover, the control fields do
not involve sudden changes in the whole process by setting
proper boundary conditions for the time derivatives of control
parameters, which means that the control fields are continuous. In addition, the performance of the NHQC+ gate is estimated in the presence of the random noise in control fields, the deviation of dipole-dipole interaction, the F\"orster defect, and decoherence based on the master equation. Both
the theoretical and numerical results show that the scheme can realize high-fidelity quantum gate.

Compared with previous schemes \cite{Qi:20,PhysRevA.94.022331,PhysRevApplied.7.054022,PhysRevA.96.052316,PhysRevLett.121.110501}, the current scheme has flexible parameter selections so that it can be compatible with many control and optimal methods. For example, in contrast to Ref. \cite{Qi:20}, the design of Hamiltonian via IBRE method in the current scheme is different from that via transitionless quantum driving (TQD) method. The design of Hamiltonian by the TQD method is to steer the dynamics along the instantaneous eigenstates of the original Hamiltonian without transitions among them. The additional Hamiltonians need to be added to compensate  the nonadiabatic errors according to the original Hamiltonian. Thus, the original Hamiltonian needs to be designed first. However, the design of Hamiltonian by the IBRE method is to reversely derive the Hamiltonian by constructing a dynamic invariant. 
In this case, the eigenvectors of the dynamic invariant as the exact evolution paths  can be obtained by designing the boundary condition of parameters. Thus, the IBRE method has more freedom to design the parameters. Accordingly, the convenience of parameter selections make the scheme well incorporate the systematic-error-sensitivity optimal control method. This enables the scheme to maintain high fidelity in the presence of errors.
As the scheme is fully compatible with the advantages of
geometric phases, reverse engineering, and
the systematic-error-sensitivity
optimal control method, it can be helpful for precise
 quantum computation in Rydberg atom systems.

\section*{Acknowledgments}
This work was supported by the National Natural Science Foundation of China under Grants
No. 11575045, No. 11874114, and No. 11674060, the Natural Science Funds for Distinguished
Young Scholar of Fujian Province under Grant 2020J06011 and Project from Fuzhou University
under Grant JG202001-2.

\appendix
\section{Demonstration of $|\tilde{\vartheta}_2(t)\rangle\langle \tilde{\vartheta}_2(t)|$ meeting the von Neumann equation}\label{appb}
From Eq.~(\ref{eq2}),  $|\psi_2(t)\rangle=\exp[i\alpha_2(t)]|\vartheta_2(t)\rangle=\exp[i\alpha_2(t)]|\tilde{\vartheta}_2(t)\rangle$ is satisfied the Schr\"odinger equation
\begin{eqnarray}
	i\frac{d|\psi_2(t)\rangle}{dt}=H_{\rm eff}|\psi_2(t)\rangle.
\end{eqnarray}
Define $\Pi_2(t)=|\tilde{\vartheta}_2(t)\rangle\langle \tilde{\vartheta}_2(t)|=|\psi_2(t)\rangle\langle\psi_2(t)|$. We can demonstrate the $\Pi_2(t)$ meets the von Neumann equation by  $|\psi_2(t)\rangle\langle\psi_2(t)|$ .
Then,
\begin{eqnarray}
	\frac{d\Pi_2(t)}{dt}&=&\frac{d|\psi_2(t)\rangle}{dt}\langle\psi_2(t)|+|\psi_2(t)\rangle\frac{d\langle \psi_2(t)|}{dt}\cr\cr
	&=&-iH_{\rm eff}|\psi_2(t)\rangle\langle\psi_2(t)|+i|\psi_2(t)\rangle\langle\psi_2(t)|H_{\rm eff}\cr\cr
	&=&-i[H_{\rm eff},|\psi_2(t)\rangle\langle \psi_2(t)|].
\end{eqnarray}
Therefore, the $|\tilde{\vartheta}_2(t)\rangle$ satisfies the von Neumann equation
\begin{eqnarray}
	\frac{d\Pi_2(t)}{dt}=-i[H_{\rm eff}, \Pi_2(t)].
\end{eqnarray}
\section{Demonstration of dynamical phase in the evolution process equal to zero}\label{appC}
In order to eliminate the dynamic phase, the evolution of the system is divided into two time intervals [0,$T/2$] and [$T/2$,$T$]. First, we calculate the time derivative of $\mu_2(t)$ by setting $\tilde{t}=T-t$ in the time interval [$T/2$,$T$] as
\begin{eqnarray}
	\frac{d\mu_2(t)}{dt}=\frac{d\tilde{t}}{dt}\frac{d}{d\tilde{t}}[-\Theta_g+\mu_2(\tilde{t})]=-\frac{d}{d\tilde{t}}\mu_2(\tilde{t}).
\end{eqnarray}
According to Eqs. (\ref{eq6}) and (\ref{gphase}) , we have
\begin{eqnarray}
	\theta_2(T)-\theta_2(T/2)&=&\int_{T/2}^{T}\frac{\dot{\mu}_2\sin^2\mu_1}{2\cos\mu_1}dt\cr\cr
	&=&\int_{T/2}^{0}\frac{\dot{\mu}_2(\tilde{t})\sin^2\mu_1(\tilde{t})}{2\cos\mu_1(\tilde{t})}d\tilde{t}\cr\cr&=&-\theta_2(T/2),\cr\cr
	\Theta_2(T)-\Theta_2(T/2)&=&\int_{T/2}^{T}\dot{\mu}_2(t)\sin^2\frac{\mu_1(t)}{2}dt\cr\cr
	&=&\int_{T/2}^{0}\dot{\mu}_2(\tilde{t})\sin^2\frac{\mu_1(\tilde{t})}{2}d\tilde{t}\cr\cr
	&=&-\Theta_2(T/2).
\end{eqnarray}
It is proved that the dynamic phase and geometric phase acquired in the time interval [0,$T/2$] is nullified by that acquired in the time interval [$T/2$,$T$].
The geometric phase is only attributed by making the value of $\mu_2$ increase $-\Theta_g$ at the moment of $t=T/2$. This can be interpreted as
\begin{eqnarray}
	\Theta_2(T/2+\Delta t)-\Theta_2(T/2)&=&-\int_{T/2}^{T/2+\Delta t}\dot{\mu}_2(t)\sin^2\frac{\mu_1(t)}{2}dt\cr\cr
	&=&\mu_2(T/2)-\mu_2(T/2+\Delta t )\cr\cr&=&\Theta_g,\cr\cr
	&&
\end{eqnarray}
when $\Delta t\rightarrow 0$. In the time interval [$T/2$, $T/2+\Delta t$], the $\mu_1(t)=\pi$. According to Eq.~(\ref{22}), the dynamic phase acquired is zero.
Thus the total dynamic phase acquired in the time interval [0, $T$] is zero.
\bibliographystyle{apsrev4-1}
\bibliography{holo}

\end{document}